\documentclass[12pt,preprint,useAMS,usenatbibi]{emulateapj}
\usepackage{color}
\usepackage{latexsym}
\usepackage{amssymb}
\usepackage{amsmath}
\usepackage{epsfig}
\usepackage{appendix}
\singlespace
\def\OVI{\hbox{O~$\scriptstyle\rm VI$}}
\def\kms{\,{\rm km\,s^{-1}}}
\def\kmsmpc{\,{\rm km\,s^{-1}\,Mpc^{-1}}}

\def\msun{\,{\rm M_\odot}}

\def\dm{\,{\rm cm^{-3}\,pc}}
\def\sfr{\,{\rm M_\odot\,yr^{-1}}}

\def\spose#1{\hbox to 0pt{#1\hss}}
\def\lta{\mathrel{\spose{\lower 3pt\hbox{$\mathchar"218$}}
     \raise 2.0pt\hbox{$\mathchar"13C$}}}
\def\gta{\mathrel{\spose{\lower 3pt\hbox{$\mathchar"218$}}
     \raise 2.0pt\hbox{$\mathchar"13E$}}}

\newcommand{\etal}{{et al.\ }}
\def\HI{\hbox{H~$\scriptstyle\rm I$}}

\lefthead{Guedes et al. 2011}
\righthead{The Eris Simulation}
\submitted{accepted by the ApJ}


\begin{document}

\title{Forming Realistic Late-Type Spirals in a $\Lambda$CDM Universe: The Eris Simulation}

\author{Javiera Guedes\altaffilmark{1,3}, Simone Callegari\altaffilmark{2}, Piero Madau\altaffilmark{1}, \& Lucio Mayer\altaffilmark{2,3}}

\altaffiltext{1}{Department of Astronomy \& Astrophysics, University of California, 1156 High Street, Santa Cruz, CA 95064}
\altaffiltext{2}{Institute for Theoretical Physics, University of Z\"urich, Winterthurerstrasse 190, CH-9057 Z\"urich, Switzerland}
\altaffiltext{3}{Institute for Astronomy, ETH Z\"urich, Wolgang-Pauli-Strasse 27, 8093 Z\"urich, Switzerland}

\begin{abstract}
Simulations of the formation of late-type spiral galaxies in a cold dark matter ($\Lambda$CDM) universe have traditionally failed to yield realistic 
candidates. Here we report a new cosmological $N$-body/smooth particle hydrodynamic (SPH) simulation of extreme dynamic range 
in which a close analog of a Milky Way disk galaxy arises naturally. Termed ``Eris", the simulation follows the assembly of a galaxy 
halo of mass $M_{\rm vir}=7.9\times 10^{11}\,\msun$ with a total of $N=18.6$ million particles (gas $+$ dark matter $+$ stars) within the final 
virial radius, and a force resolution of 120 pc. It includes radiative cooling, heating from a cosmic UV field and supernova 
explosions (blastwave feedback), a star formation recipe based on a high gas density threshold ($n_{\rm SF}=5$ atoms cm$^{-3}$ rather than 
the canonical $n_{\rm SF}=0.1$ atoms cm$^{-3}$), and neglects any feedback from an active galactic nucleus. Artificial images are generated to 
correctly compare simulations with observations. At the present epoch, the 
simulated galaxy has an extended rotationally-supported disk with a radial scale length $R_d=$ 2.5 kpc, a gently falling rotation 
curve with circular velocity at 2.2 disk scale lengths of $V_{2.2}=214\,\kms$, an $i$-band bulge-to-disk ratio $B/D=0.35$, and a baryonic 
mass fraction within the virial radius that is 30\% below the cosmic value. The disk is thin, has a typical \HI-to-stellar mass ratio,
is forming stars in the region of the $\Sigma_{\rm SFR}$-$\Sigma_{\rm HI}$ plane occupied by spiral galaxies, and falls on the 
photometric Tully-Fisher and the stellar mass-halo virial mass relations. Hot ($T>3\times 10^5$ K), X-ray luminous halo gas makes only 
26\% of the universal baryon fraction and 
follows a ``flattened" density profile $\propto r^{-1.13}$ out to $r=100$ kpc. Eris appears then to be the first cosmological hydrodynamic simulation 
in which the galaxy structural properties, the mass budget in the various components, and the scaling relations between mass and luminosity
are all consistent with a host of observational constraints.  A twin simulation with a low star formation density threshold results in a galaxy with a 
more massive bulge and a much steeper rotation curve, as in previously published work. A high star formation threshold appears therefore key in obtaining 
realistic late-type galaxies, as it enables the development of an inhomogeneous interstellar medium where star formation and heating
by supernovae occur in a clustered fashion. The resulting outflows at high redshifts reduce the baryonic content of galaxies and preferentially 
remove low angular momentum gas, decreasing the mass of the bulge component. Simulations of even higher resolution that follow the assembly of 
galaxies with different merger histories shall be used to verify our results.

\end{abstract}


\section{Introduction}

The formation of realistic late-type spirals has been a long standing problem of galaxy formation in a $\Lambda$CDM 
universe. Within this framework, baryons condense at the center of dark matter halos and acquire angular momentum through tidal torques from 
nearby structures \citep{fall80}. A centrifugally-supported baryonic disk forms, with a size that depends on the fraction of the 
original angular momentum that is retained during the contraction. In numerical simulations of this process, however, a fundamental 
``angular momentum problem" arises, as galaxies are produced with a baryonic component that is quite deficient 
in angular momentum compared to real spirals \citep{navarro91,navarro00}. Aside from artificial losses of angular momentum caused by insufficient resolution and other numerical 
effects \citep{okamoto03,governato04,kaufmann07}, this failure has traditionally been traced back 
to the very nature of the hierarchical buildup of structures: dynamical friction transfers the orbital angular momentum of merging substructures 
to the outer halo, and causes the associated cold baryons to sink to the center of the proto-galaxy and form a spheroid rather than a disk 
\citep[e.g.][]{maller02}. 

\begin{deluxetable*}{lccccccccccccccr}
\tablecaption{Properties of the simulated galaxy \label{simsum}}
\tablewidth{0pt}
\tablehead{\colhead{Galaxy} & \colhead{M$_{\rm vir}$} & \colhead{$V_{\rm peak}$} & \colhead {M$_{*}$} & \colhead{$f_b$} & \colhead{$f_{\rm cold}$} & 
\colhead{$m_{\rm DM}$} &  \colhead{$m_{\rm SPH}$} & \colhead{$\epsilon_G$} & \colhead{$\epsilon_{\rm SF}$} & \colhead{$N_{\rm DM}$} & \colhead{$N_{\rm gas}$} & 
\colhead{$N_*$} & \colhead{$B/D$} & \colhead{$R_d$}\\
}
\tabletypesize{\footnotesize}
\startdata
Eris ($z=0$) & 7.9 & 238 & 3.9 & 0.121 & 0.12 & 9.8 & 2 & 120 & 0.1 & $7.0$ & $3.0$ & $8.6$ & 0.35 & 2.5\\
Eris ($z=1$) & 5.4 & 237 & 2.9 & 0.126 & 0.40 & 9.8 & 2 & 120 & 0.1 & 4.8 & 2.0 & 6.2 & 0.30 & 1.8\\
ErisLT ($z=1$) & 5.5 & 308 & 3.4 & 0.158 & 0.18 & 9.8 & 2 & 120 & 0.05 & 4.9 & 2.9 & 8.3 & 0.42 & 1.4
\enddata
\tablecomments{Columns 2, 3, 4, 5, and 6 list the virial mass (in units of $10^{11}\,\msun$), peak circular velocity (in $\kms$), 
total stellar mass of the halo (in units of $10^{10}\,\msun$), baryonic mass fraction, and cold ($T<3\times 10^4$ K) gas fraction. 
Columns 7 and 8 list the mass resolution of individual dark matter and SPH particles (in units of $10^4\,\msun$), and columns 
9 and 10 the spline gravitational force softening (in pc) and the star formation efficiency. Columns 11, 12, and 13 list 
the total number (in units of $10^6$) of dark matter, gas, and star particles within the virial radius of the halo. Columns 
14 and 15 list the bulge-to-disk ratio and disk scale length (in kpc) estimated from the $i$-band photometric decomposition.}
\\
\end{deluxetable*}

A popular solution to the angular momentum problem envisions energy injection from supernovae (SNe) and evolving stars as a mechanism to 
prevent efficient gas cooling and condensation and to remove low angular momentum material from the central part of galaxies. Modern simulations 
with improved resolution and more effective recipes for SN feedback \citep{robertson04,governato07,scannapieco09,stinson10,piontek11,brooks11} 
have yielded rotationally-supported disks with realistic exponential scale lengths, not only in galaxies formed in relative isolation but also in those that
are accreted by massive groups with a dominant central elliptical \citep{feldmann10a,feldmann10b}. They have also modified the standard picture of gas 
accretion and cooling onto galaxy disks: for galaxies up to Milky Way masses, gas acquired through filamentary ``cold flows" that was 
never shock-heated to the halo virial temperature is largely responsible for star formation in the disk at all times \citep{brooks09,keres09,ceverino10}. 
Yet, these simulations typically continue to produce centrally-concentrated systems, with rotation curves that rise steeply towards the center: 
simulated disk galaxies 
fall exclusively in the S0 or Sa category, leaving late-type spirals with negligible bulges and large disks with flat rotation curves -- such as our 
own Milky Way -- as an unsolved puzzle. Two recent exceptions are the simulations of \citet{agertz11} and \citet{governato10}. 
In the first, replicas of Sb/Sc galaxies with moderate bulges were obtained with a low efficiency of star formation that may implicitly mimic 
the bottleneck of the conversion of atomic gas into molecular, at the expense of producing stellar disks that are much more massive 
than expected at a given halo mass \citep[e.g.][]{guo10}. In the second, a realistic bulgeless dwarf galaxy with a shallow 
central dark-matter profile was generated by resolving the inhomogenous interstellar medium (ISM) and the process of energy injection from multiple 
SNe in clustered star forming regions. In this paper we extend the latter approach to massive galaxy scales, and 
present initial results from a new SPH cosmological simulation 
of high dynamic range that includes radiative cooling, heating from a cosmic UV field, SN feedback, and a star 
formation recipe based on a high gas density threshold as in \citet{governato10}. It is this last feature, we argue, that is key to the 
formation of a more realistic massive late-type spiral in $\Lambda$CDM.

\section{Simulation setup}

Dubbed ``Eris", the simulation described in this paper is part of a campaign of extreme resolution simulations of the formation of
Milky Way-sized galaxies \citep{diemand07,diemand08}. It was performed in a {\it Wilkinson Microwave Anisotropy Probe} 3-year cosmology, 
$\Omega_M=0.24$, $\Omega_\Lambda=1-\Omega_M$, $\Omega_b=0.042$, $H_0=73\,\kmsmpc$, $n=0.96$, $\sigma_8=0.76$, running the 
parallel, spatially and temporally adaptive, treeSPH-code {\it GASOLINE} \citep{wadsley04} for 1.5 milion cpu hours. 
The target halo was identified at $z=0$ in a 
low-resolution, dark matter-only, periodic box of 90 Mpc on a side. It was choosen to have a similar mass as the Milky Way and a rather 
quiet late merging history, i.e. to have had no major mergers (defined as mass ratio $\ge 1/10$) after $z=3$. New initial conditions were then 
generated with improved mass resolution, centered around a Lagrangian sub-region of 1 Mpc on a side, using the standard ``zoom-in" technique to 
add small-scale perturbations. High-resolution particles were further split into 13 million dark matter particles and an equal number of 
gas particles, for a final dark and gas particle mass of 
$m_{\rm DM}=9.8\times 10^4\,\msun$ and $m_{\rm SPH}=2\times 10^4\,\msun$, respectively. The gravitational softening length, $\epsilon_G$, 
was fixed to 120 physical pc for all particle species from $z=9$ to the present, and evolved as $1/(1+z)$ from $z=9$ to the 
starting redshift of $z=90$.

The version of the code used in this study includes Compton cooling, atomic cooling, and metallicity-dependent radiative 
cooling at low temperatures \citep{mashchenko06}. A uniform UV background modifies 
the ionization and excitation state of the gas and is implemented using a modified version of the \citet{haardt96} spectrum. 
Three parameters characterize the star formation and feedback recipes: (a) the star formation threshold $n_{\rm SF}$, (b) the 
star formation efficiency $\epsilon_{\rm SF}$, and (c) the fraction of SN energy that couples to the ISM $\epsilon_{\rm SN}$.      
Star formation occurs when cold ($T<3\times 10^4$ K), virialized gas reaches a threshold density $n_{\rm SF}=5$ atoms cm$^{-3}$ and is part 
of a converging flow. It proceeds at a rate 
\begin{equation}
d\rho_*/dt=\epsilon_{\rm SF} \rho_{\rm gas}/t_{\rm dyn} \propto \rho_{\rm gas}^{1.5}
\label{eq:KS}
\end{equation}
(i.e. locally enforcing a Schmidt law), where $\rho_*$ and $\rho_{\rm gas}$ are the stellar and gas densities, and $t_{\rm dyn}$ is the 
local dynamical time. We choose $\epsilon_{\rm SF}=0.1$. An additional identical run with $\epsilon_{\rm SF}=0.05$, the value
adopted in \citet{governato07} and \citet{brook10}, yielded a galaxy with nearly identical structural properties, and will
be discussed in a forthcoming paper. Each star particle is created stochastically with an initial mass $m_*=6\times 10^3\,\msun$, 
and the gas particle that spawns the new star has its own mass reduced accordingly. A star particle represents a simple stellar population 
with its own age, metallicity, and a \citet{kroupa01} initial stellar mass function (IMF). Each SN deposits metals and a net energy of 
$\epsilon_{\rm SN} \times 10^{51}\,$ergs into the nearest neighbor gas particles, with $\epsilon_{\rm SN}=0.8$ (the same value adopted in 
previous simulations). The heated gas has its cooling shut off until the end of the snowplow phase of the SN blastwave, which is set by the 
local gas density and temperature and by the total amount of energy injected $E$ \citep{stinson06}. For the typical ISM conditions at 
threshold found in this study, this translates into regions of size $R_E\sim 30 E_{51}^{0.32}$ pc heated by individual SNe and having 
their cooling shut off for a timescale $t_E\sim 5\times 10^5 E_{51}^{0.31}$ yr, where $E_{51}\equiv E/10^{51}\,{\rm ergs}$. The energy 
injected by many SNe adds up to create larger hot bubbles and longer shutoff times. No feedback from an active galactic nucleus was included.

\begin{figure*}[th]
\centering
\includegraphics[width=0.48\textwidth]{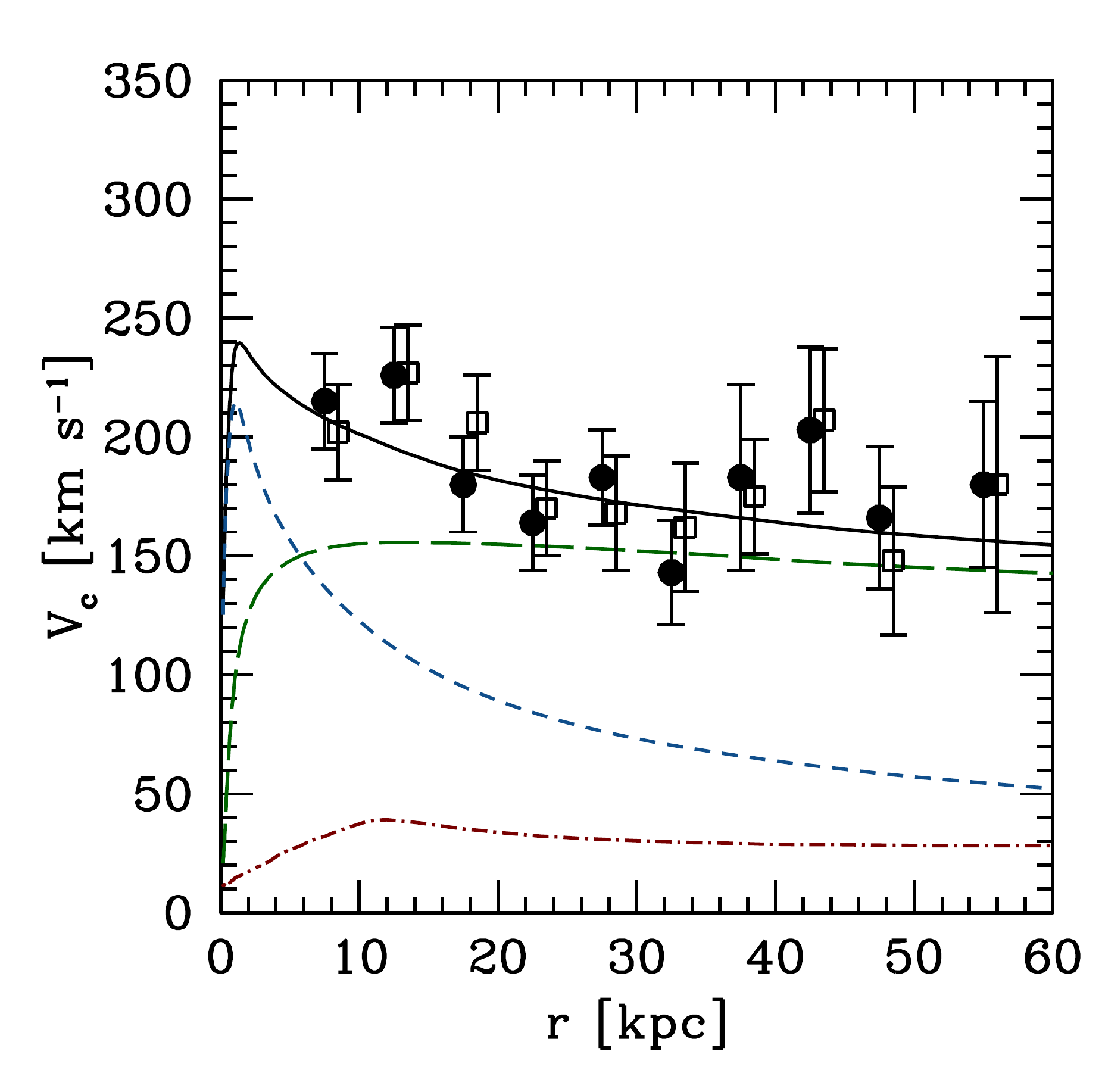}
\includegraphics[width=0.48\textwidth]{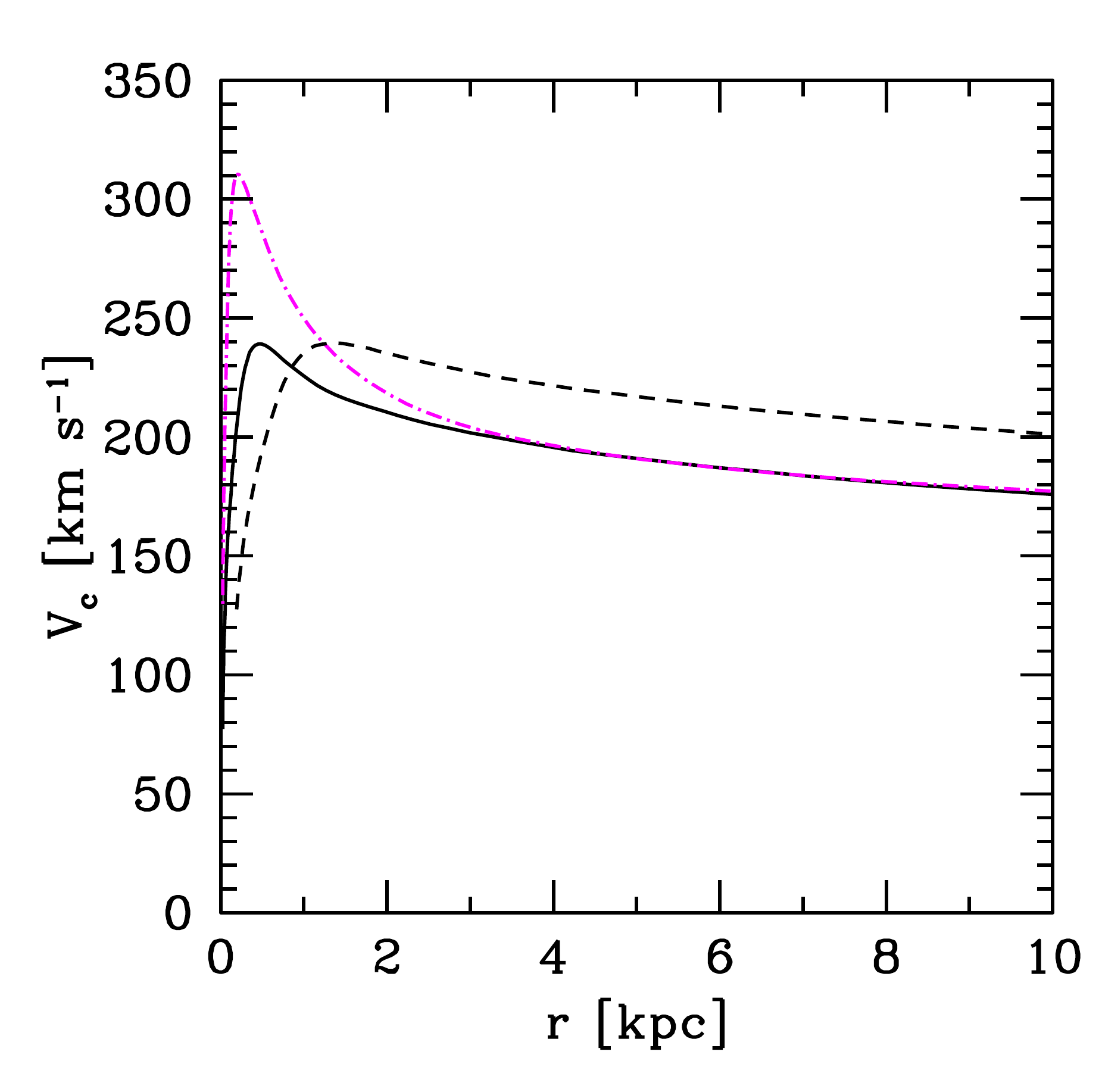}
\caption{\footnotesize {\it Left panel}: The rotation curve of the simulated Milky Way-sized galaxy (``Eris") at $z=0$.
The figure shows the contributions to the circular velocity $V_c=\sqrt{GM(<r)/r}$ of the various mass components: dark matter ({\it 
long-dashed curve}), stars ({\it short-dashed curve}), gas ({\it dot-short dashed curve}), and total ({\it solid curve}). The data points 
show two realizations of the rotation curve of the Milky Way from observations of blue horizontal-branch halo stars in the 
{\it Sloan Digital Sky Survey} \citep{xue08}, and have been offset slightly from each other in radius for clarity. 
{\it Right panel}: The total inner rotation curve at $z=1$ for the fiducial high-threshold simulation (Eris, {\it solid 
line}) and for the low-threshold (ErisLT, {\it dot-short dashed line}) twin run. The short-dashed line shows Eris' inner rotation curve 
at $z=0$ for comparison. The star formation threshold has a significant effect on the mass distribution: a more prominent stellar 
bulge forms at early times in ErisLT and is responsible for the peaked rotation curve. 
}
\label{fig1}
\vspace{+0.3cm}
\end{figure*}

The adoption of a density threshold for star formation that is 50 times higher than in many previous lower-resolution studies is possible 
owing to the high mass and spatial resolution of this run, which resolves the giant cloud complexes where star formation actually occurs in the ISM
and the true scale height of the neutral atomic ISM. In particular, the local Jeans length corresponding to our density threshold (for $T=10^3$ K, 
a lower bound on the typical temperature of the cold gas in the simulations) is resolved with more than 5 SPH smoothing lengths, thus preventing artificial 
fragmentation \citep{bate97}. While not as high as the value of $n_{\rm SF}=100$ atoms cm$^{-3}$ used in \citet{governato10} dwarf galaxy simulation, whose
particle mass was substantially lower and allowed to resolve the Jeans length of star forming gas at much higher densities,
our star formation threshold is still large enough to allow the development of a clumpy, inhomogeneous ISM with more localized
energy injection by multiple overlapping SN explosions. This allows galactic pressure-driven outflows 
to develop and remove low-angular momentum material. To demonstrate the important role of the star formation threshold on the structural 
properties of massive galaxies, we have run a low-threshold twin simulation (termed ``ErisLT") with $n_{\rm SF}=0.1$ atoms cm$^{-3}$. We have kept all the 
other simulation parameters fixed (same mass and spatial resolution and identical feedback scheme) except for the star formation efficiency parameter, 
$\epsilon_{\rm SF}$, which was lowered from 0.1 (Eris) to 0.05 (ErisLT) to match the observed normalization of the star formation density
in local galaxies \citep[see][]{governato10}. ErisLT was stopped at redshift 0.7 in order to limit the computational burden.

\begin{figure*}[th]
\centering
\includegraphics[width=.86\textwidth]{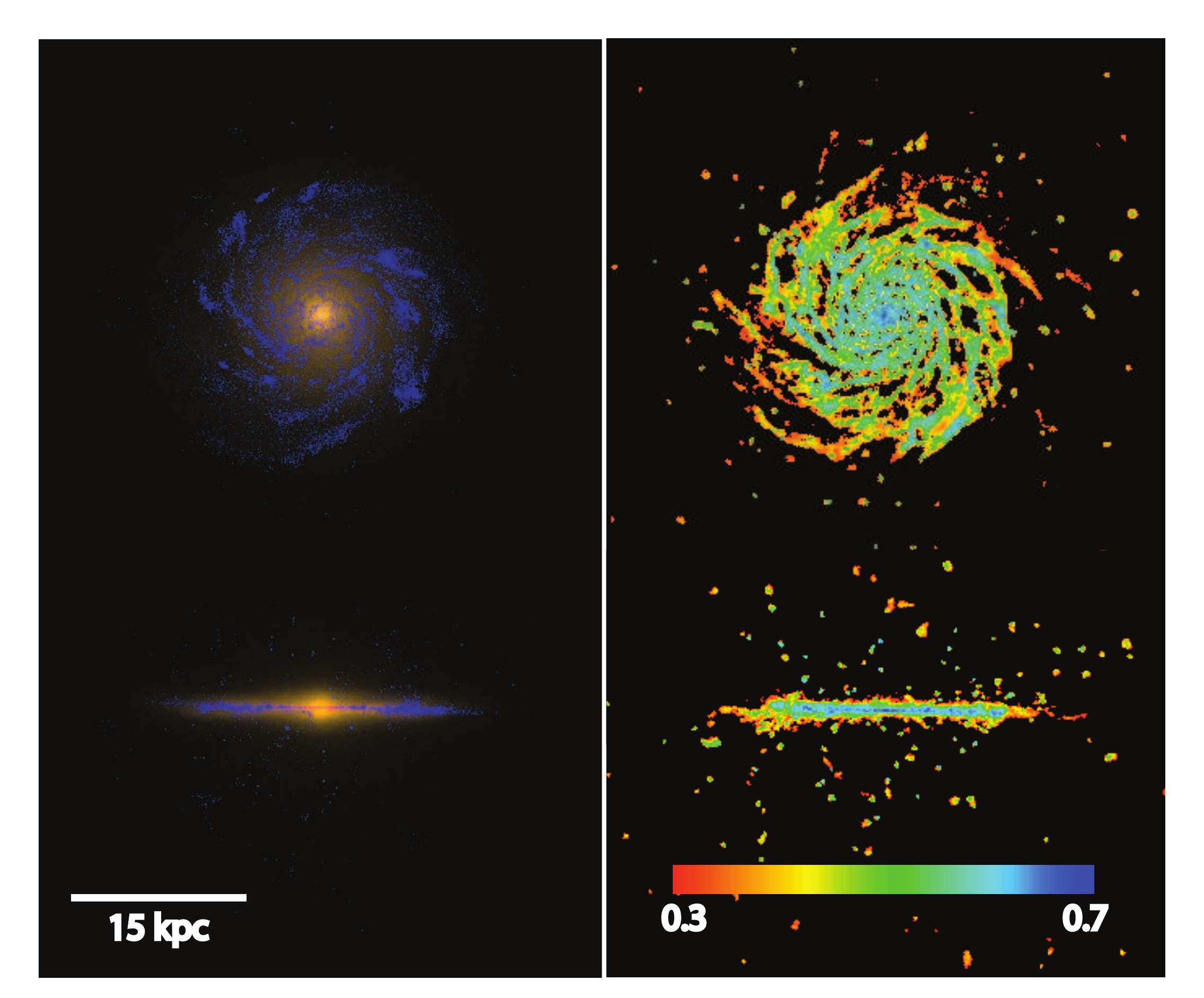}
\vspace{+0.3cm}
\caption{{\it Left panel:} The optical/UV stellar properties of Eris at $z=0$. The images, created with the radiative transfer code 
{\sc Sunrise} \citep{jonsson06}, show an $i$, $V$, and $FUV$ stellar composite of the simulated galaxy seen face-on and edge-on. A Kroupa IMF
was assumed. {\it Right panel:} Projected face-on and edge-on surface density maps of Eris's neutral gas at $z=0$. The color bar shows the neutral 
gas fraction. 
}
\label{fig2}
\vspace{+0.4cm}
\end{figure*}

\section{Results}

In this first paper we focus on the final $z=0$ state and properties of the galaxy and on the comparison with observational constraints. 
The main characteristics of the simulated galaxy are listed in Table 1.

\subsection{Structural Parameters}
At the present epoch, Eris is a massive, barred, late-type spiral with structural properties consistent with those of Sb/Sbc galaxies, 
of which our own Milky Way is an example. It has a virial radius of $R_{\rm vir}=239$ kpc (defined as the radius enclosing 
a mean density of $93\rho_{\rm crit}$, \citealt{bryan98}), total mass $M_{\rm vir}=7.9\times 10^{11}\,\msun$,
spin parameter $\lambda=0.019$ (\`{a} la \citealt{bullock01}), and 7.0, 3.0, and 8.6 million dark matter, gas, and star particles within $R_{\rm vir}$, 
respectively. The minimum smoothing length for gas particles is 5 times smaller than the force softening. The total mass enclosed within 60 kpc 
is $M_{<60}=3.3\times 10^{11}\,\msun$. The rotation curve, shown in the left panel of Figure \ref{fig1}, has a peak circular velocity of 
$V_{\rm peak}=238\,\kms$ (reached at 1.34 kpc) and a value at 8 kpc (the solar circle) of $V_{c,\odot}=206\,\kms$. 
Its overall shape out to 20 kpc, including the peak 
in the central bulge-dominated kpc, is reminiscent of the recent reconstruction of the Milky Way rotation curve by \citet{sofue09}.
The circular velocity decreases gently to distances of 60 kpc from its value at the solar radius, in agreement with 
observations of blue horizontal-branch halo stars in the {\it Sloan Digital Sky Survey} \citep{xue08}. The measured $V_{c,\odot}, 
M_{<60}$, and $M_{\rm vir}$ agree within the errors with the values of $V_{c,\odot}=221\pm 18\,\kms$, $M_{<60}=4.0\pm 0.7\times 10^{11}\,\msun$,
and $M_{\rm vir}=1.0^{+0.3}_{-0.2} \times 10^{12}\,\msun$ derived recently for the Milky Way using the narrow GD-1 stream of stars 
\citep[for $V_{c,\odot}$,][]{koposov10} and halo stars as kinematic tracers \citep[for $M_{<60}$ and $M_{\rm vir}$,][]{xue08}.

\subsection{Brightness Profile}
\begin{figure}[t]
\centering
\includegraphics[width=.45\textwidth]{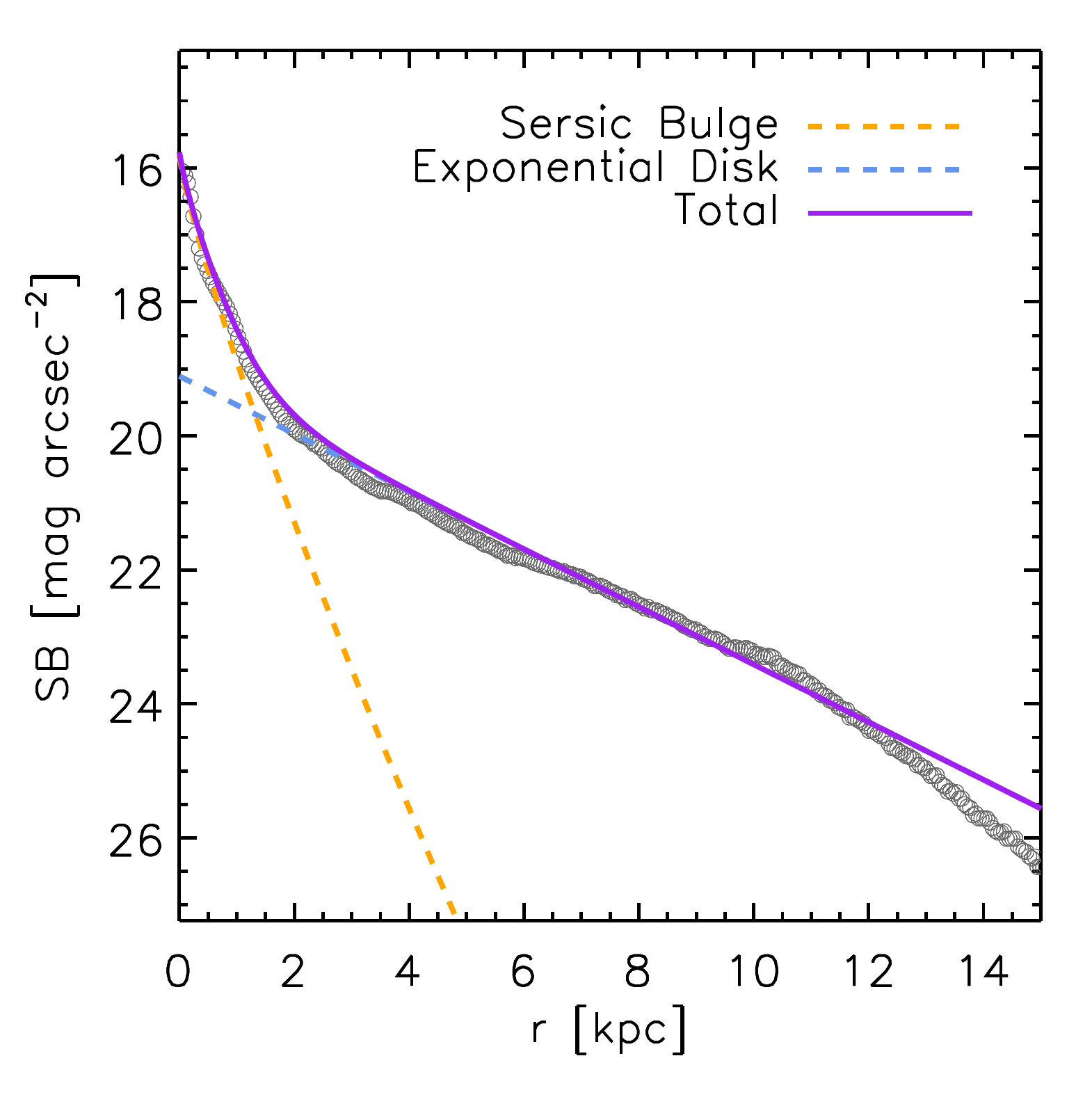}
\caption{The 1D $i$-band radial surface brightness profile of Eris at $z=0$. This is well fitted by a S\'{e}rsic bulge with index $n_s=1.4$,
an exponential disk with scale length $R_d=2.5$ kpc, and a bulge-to-disk ratio $B/D=0.35$. The dust reddened, face-on 2D light distribution created
by {\sc Sunrise}  was analyzed with {\sc Galfit} \citep{peng02} following a procedure similar to that detailed in \citet{weinzirl09}.
The ``downbending" in the brightness exponential profile at about 5 disk scale length and the surface brightness where the break 
occurs, 23.5 $i$-mag arcsec$^{-2}$, are characteristic of late-type spiral galaxies \citep{pohlen06}. 
}
\label{fig3}
\vspace{+0.3cm}
\end{figure}

To correctly compare simulations with observations we created artificial images of our simulations and from them measured photometric 
bulge-to-disk ratios and disk scale lengths. The mock images were created using the radiation transfer code {\sc Sunrise} \citep{jonsson06}, 
which produces spectral energy distributions using the age
and metallicities of each simulated star particle, and takes into account the three-dimensional effect of dust reprocessing. The results 
for a Kroupa IMF are shown in Figure \ref{fig2}. A 2D photometric decomposition
was performed on the dust-reddened $i$-band light distribution with the {\sc Galfit} program \citep{peng02}. At the present epoch,
the total $i$-band magnitude is $M_i=-21.7$, and a stellar disk with a scale length $R_d=2.5$ kpc dominates the light distribution (Fig. \ref{fig3}).
The disk scale length is comparable to the value $R_d=2.3\pm 0.6$ kpc, adopted for the Milky Way in the compilation by \citet{hammer07}, 
and with the scale length of the Milky Way thin disk, $2.6$ kpc, as traced by M dwarfs in the solar neighborhood \citep{juric08}.
Its value also agrees with the scaling relations of spiral galaxies \citep{courteau07}. The SDSS $u-g=1.03$ mag and $g-r=0.49$ mag integrated 
colors, obtained directly from the {\sc Sunrise} images, fall within 1$\sigma$ of the mean optical colors of late-type galaxies as luminous as Eris 
\citep{blanton03}. The ``downbending" observed in Eris' brightness exponential profile at about 4 disk scale lengths appears to be characteristic of 
late-type spirals \citep{pohlen06}. As in the sample of truncated late-type spirals of \citet{bakos08}, there is no break 
in the stellar surface mass density profile of Eris: rather, Eris' stellar age profile shows a ``U shape" with a minimum of 6 Gyr at the break radius, 
explaining the origin of the break as a radial change in stellar population likely caused by the stochastic radial migration of young stars from
the inner parts of the disk to the outskirts \citep{roskar08}.

Eris' bulge-to-disk ratio (as determined by a two-component fit to the $i$-band surface brightness profile), $B/D=0.35$,
is also typical of Sb spirals, which are characterized by a median ($\pm 68/2$ per cent) value $\log B/D=-0.53^{+0.27}_{-0.30}$,
and of many Sbc galaxies, which have $\log B/D=-0.86^{+0.34}_{-0.40}$  \citep{graham08}. A three-component decomposition (disk$+$bar$+$bulge)
will lower the $B/D$ ratio further. The bulge S\'{e}rsic index, $n_s=1.4$, is indicative of a ``pseudobulge" rather than a classical one:
according to \citet{weinzirl09}, $\sim 3/4$ of all bright spirals have low $n_s\le 2$ bulges. Eris' large final disk (disk-to-total ratio $D/T=0.74$) 
is not typically found in lower-resolution simulations of Milky Way-sized galaxies that impose no restrictions on merger history: e.g., only one of 
the eight galaxies simulated by \citet{scannapieco10} has a photometric $D/T$
as large as 0.68 (and six have $D/T<0.5$), and only one out of the six galaxies above $M_{\rm vir}=10^{11}\,\msun$
simulated by \citet{brooks11} has a disk-to-total ratio comparable to Eris' (``h239", which is offset, however, from the stellar mass-halo mass relation).

\subsection{Stellar Content}
Eris' total mass in baryons is $M_b=9.5\times 10^{10}\,\msun$, corresponding to a mass fraction $f_b=0.12$ that is
30\% lower than the universal value (for the adopted cosmology) of 0.175. Stars (and their remnants) comprise 41\% of all baryons 
within $R_{\rm vir}$: the total stellar mass, $M_*=3.9\times 10^{10}\,\msun$, is comparable to the value estimated for the 
Milky Way, $4.9-5.5\times 10^{10}\,\msun$, by \citet{flynn06}. 

\begin{figure*}[th]
\centering
\includegraphics[width=0.46\textwidth]{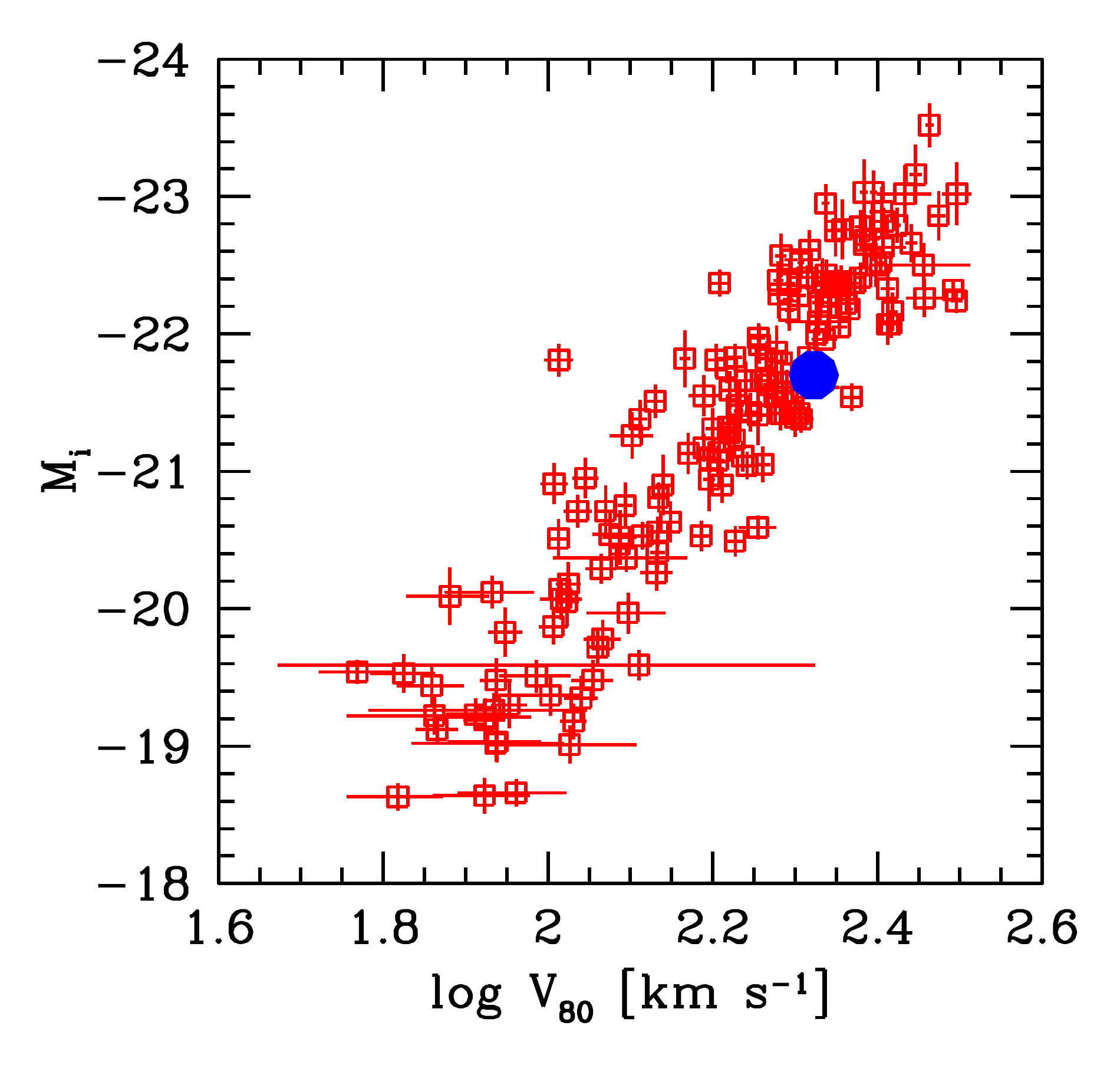}
\includegraphics[width=0.46\textwidth]{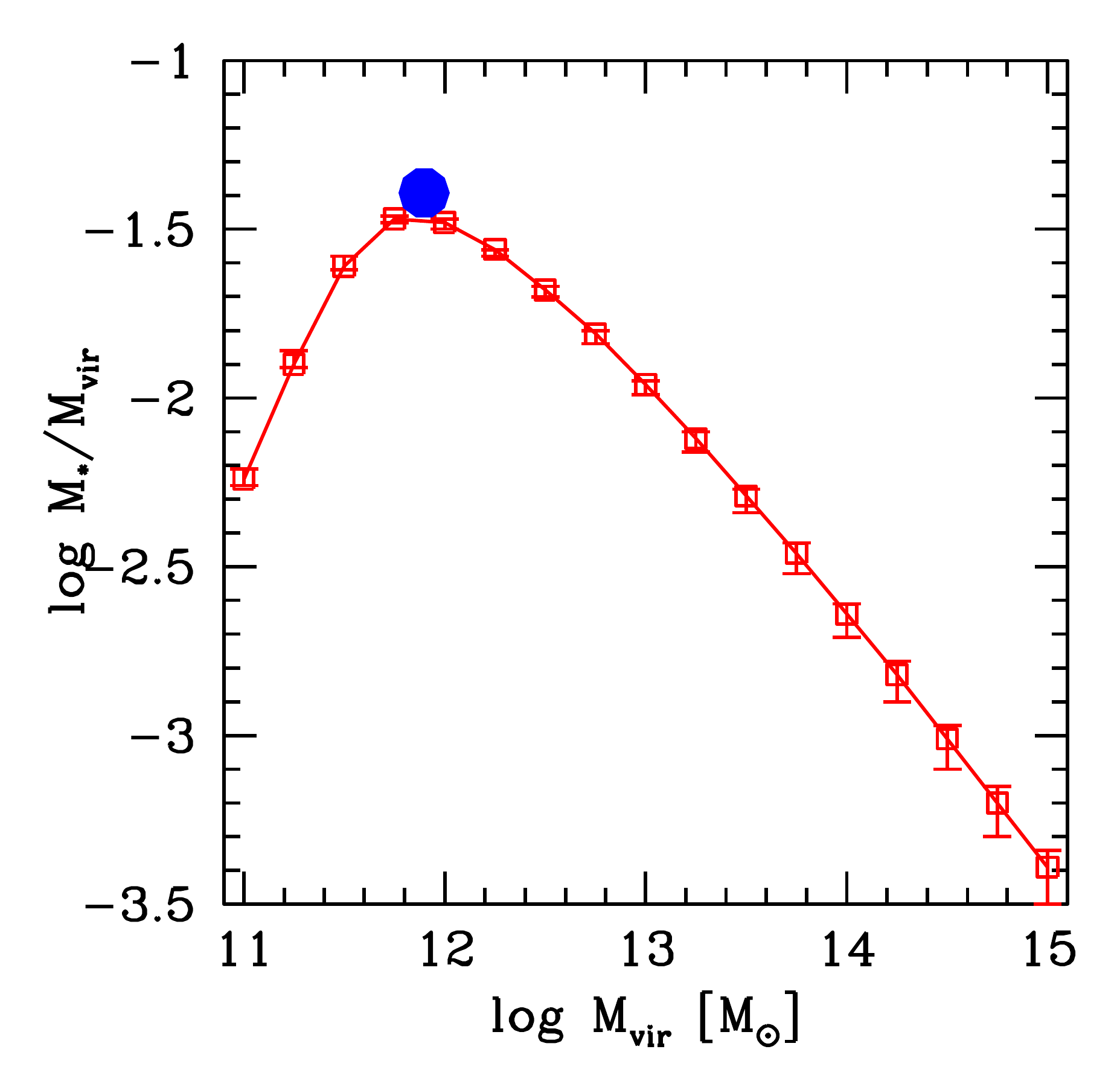}
\caption{\footnotesize {\it Left panel:} The $i$-band Tully-Fisher relation for the \citet{pizagno07} galaxy sample 
({\it empty squares with error bars}). {\it Filled circle:} The Eris simulation. Here $V_{80}$ denotes the 
circular velocity at the radius containing 80\% of the $i$-band flux, as defined by \citet{pizagno07}.   
{\it Right panel:} The stellar mass - halo mass relation at $z=0.1$ from \citet{behroozi10}, modified for a Kroupa IMF ({\it empty squares 
with error bars}). Errors bars include only statistical uncertainties. {\it Filled circle:} The Eris simulation with a 
photometric stellar mass of ${\cal M}_*=3.2\times 10^{10}\,\msun$ and a virial mass of $M_{\rm vir}=7.9\times 10^{11}\,\msun$
(see text for details).
}
\label{fig4}
\vspace{+0.3cm}
\end{figure*}

To make a bias-free comparison with the stellar mass-halo mass relation derived from the abundance matching technique by 
\citet{behroozi10} we adopt the following procedure. We fit the SDSS $u,g,r,i,z$ broadband colors from the mock {\sc Sunrise} images     
with the flexible stellar population synthesis code of \citet{conroy09}: the fit assumes a Kroupa IMF and provides a {\it 
photometric} stellar mass estimate of ${\cal M}_*=3.2\times 10^{10}\,\msun$ (Conroy, private communication), 18\% lower
than the value directly measured in the simulation. The photometric stellar mass of Eris can now be weighted self-consistently against 
the \citet{behroozi10} average stellar mass-halo relation (which uses a \citealt{chabrier03} IMF), free of IMF systematics, after offsetting 
all \citet{behroozi10} stellar masses by 0.06 dex (to correct from Chabrier to Kroupa IMF).  The comparison, depicted in the right panel of Figure \ref{fig4}, 
demonstrates that Eris' implied ``baryon conversion efficiency", $\eta\equiv ({\cal M}_*/M_{\rm vir}) \times (\Omega_M/\Omega_b)=23\%$, is in excellent 
agreement with that predicted by the abundance matching technique. This contrasts with the recent analysis of many hydrodynamic 
simulations of galaxy formation by \citet{guo10}, who show that the great majority of them lock too many baryons into 
stars to be viable models for the bulk of the observed galaxy population. Note that the intrinsic scatter in the stellar
mass at a given halo mass is estimated to be 0.17 dex, independent of halo mass \citep{yang09}.

With a circular velocity at the radius, $R_{80}=6.8$ kpc, containing 80\% of the $i$-band flux of $V_{80}=210\,\kms$, our galaxy lies
close to the Tully-Fisher relation of the \citet{pizagno07} galaxy sample (see the left panel of Fig. \ref{fig4}).
As discussed in \citet{pizagno07}, the Tully-Fisher uses $V_{80}$ as the primary velocity measure rather than $V_{2.2}$, the
circular velocity at 2.2 disk scale lengths, since the former is less sensitive to the degeneracies of bulge-disk decomposition.
The ratio $V_{2.2}/V_{200}$=$214\,\kms/129\,\kms=1.66$ in Eris, where
$V_{200}$ is the circular velocity at the radius enclosing a mean overdensity of $200\,\rho_{\rm crit}$ ($R_{200}=177$ kpc), is
equal to the value suggested by the dynamical model for the Milky Way of \citet{klypin02}. It is also consistent 
with the recent measurements of the virial mass of the Milky Way by \citet{smith07} and \citet{xue08}, implying
$V_{2.2}/V_{200} =1.48^{+0.25}_{-0.26}$ and $V_{2.2}/V_{200} =1.67^{+0.31}_{-0.24}$, respectively.\footnote{The $V_{2.2}/V_{200}$
ratios from \citet{smith07} and \citet{xue08}, were computed by \citet{dutton10} from these data sets after converting
different virial mass definitions and for an assumed Milky Way's $V_{2.2}=220\,\kms$.}\, Note that while there is an unsettled 
disagreement among estimates of the Galaxy's virial mass\footnote{Recent estimates of the virial mass of the Milky Way range from
 $M_{\rm vir}=1.0^{+0.3}_{-0.2} \times 10^{12}\,M_{\odot}$ from blue horizontal branch kinematics \citep{xue08} to $1.2^{+0.7}_{-0.4} \times 10^{12}\,M_{\odot}$ from studies based on the properties of the Large and Small Magellanic Clouds \citep{busha11}.}, Eris is 
likely 20\% less massive than the Milky Way and therefore the agreement 
between simulated and observed kinematic data should be confirmed in future simulations of slightly more massive halos and 
different accretion histories. Like the Milky Way, however, Eris is offset relative to determinations using various dark halo mass 
tracers for late-type disk galaxies by \citet{dutton10}, who predict for typical dark matter halos with $V_{2.2}=220\,\kms$ the ratio 
$V_{2.2}/V_{200}=1.11^{+0.22}_{-0.20}~(2\sigma)$.

\subsection{Gas Content}
Eris' \HI\ mass is $M_{\rm HI}=1.9\times 10^9\,\msun$, comparable to the \HI\ mass 
estimated for the Milky Way disk by \citet{nakanishi03}, but smaller than the value of $\sim 5\times 10^9\,\msun$ given 
by \citet{wolfire03}. The \HI-to-stellar mass ratio, $1.9\times 10^9\,\msun/3.9\times 10^{10}\,\msun=0.049$, is 
equal to the median value observed in the GASS survey \citep{catinella10} for galaxies of comparable stellar mass. Eris' \HI\ 
disk extends out to about 15 kpc (6 stellar disk scale lengths), similar to the size of the \HI\ disk of the Milky Way \citep{nakanishi03}. 
Clustered SN explosions create a large number of holes in the face-on \HI\ distribution of Eris (Fig. \ref{fig2}) due to 
bubbles of hot gas expanding perpendicular to the disk. These holes are mostly located within the bright optical disk and preferentially in 
regions of high star formation: they are kpc in size, as observed, e.g., in the nearby low-inclination spiral galaxy NGC 6946 \citep{boomsma08}.  

\begin{figure}[th]
\centering
\includegraphics[width=.47\textwidth]{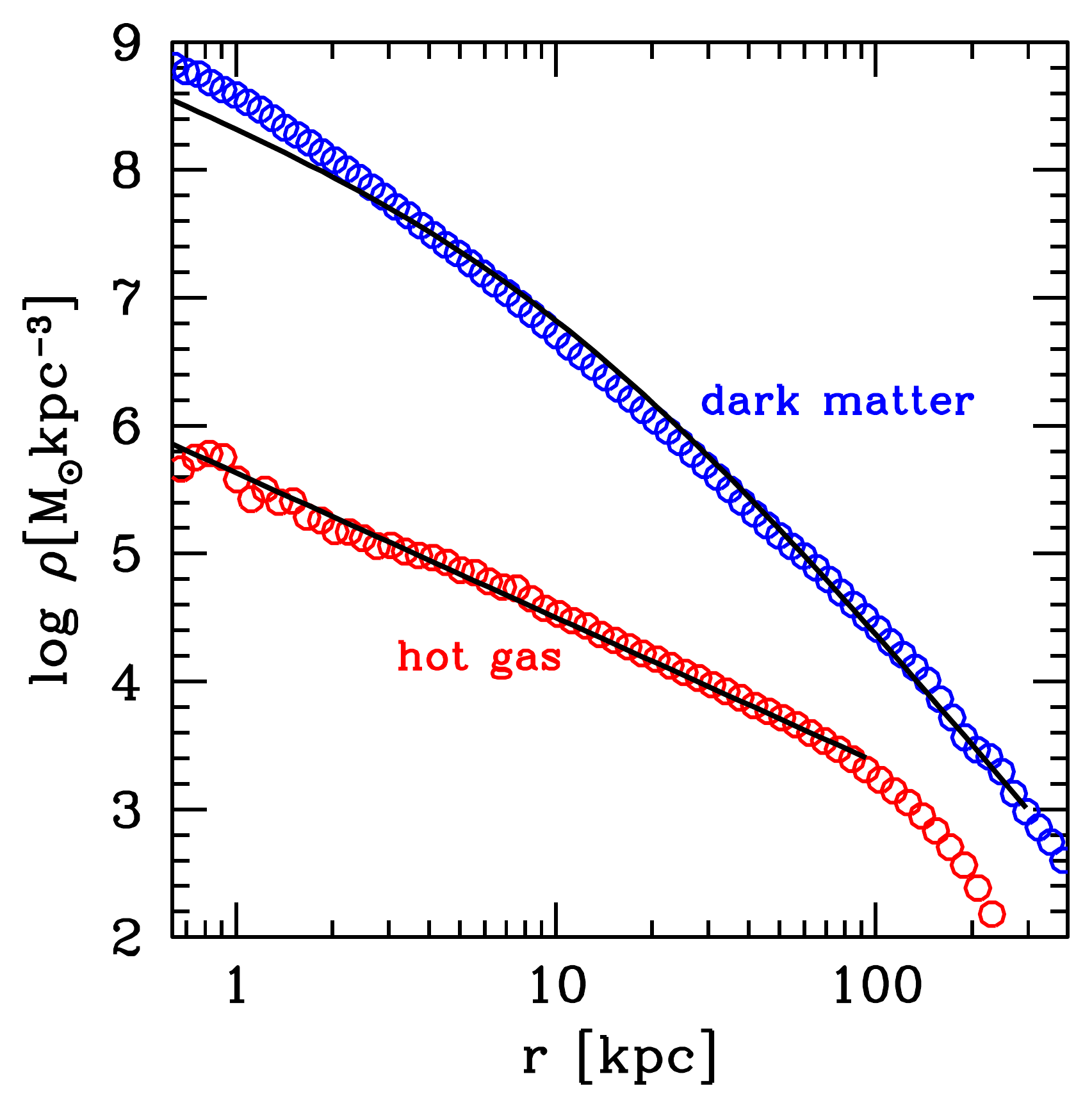}
\caption{The average dark matter ({\it blue empty dots}) and hot ($T>3\times 10^5\,$K) gas ({\it red empty dots}) density profiles of Eris at $z=0$. The
solid lines show the best-fit NFW profile for the dark matter ({\it upper curve}) and the best-fit power-law profile (with slope $-1.13$) for the 
hot gas ({\it lower curve}). The best-fit NFW profile is characterized by a large halo concentration parameter $c\equiv R_{\rm vir}/R_s=22$ as the 
dark matter halo contracts in response to the condensation of baryons in its center.
}
\label{fig5}
\vspace{+0.cm}
\end{figure}

About $6.7\times 10^9\,\msun$ are found in a cold phase below $T=3\times 10^4$ K. This is comparable to the total mass of the atomic and
warm ionized medium inferred for the Milky Way \citep[e.g.][]{ferriere01}.    
The gas mass that is hot ($T>3\times 10^5\,$K) and thus potentially X-ray luminous is $M_X=3.6\times 10^{10}\,\msun$, 
63\% of the total gas content. For comparison, 12\% of the gas is in the cold phase and 25\% is in a warm phase with
$3\times 10^4~{\rm K} <T< 3\times 10^5$ K. The fractions of cold, warm, and hot gas within its inner 20 kpc are 83.5\%, 1.5\%, and 15\%, respectively.  
Hot gas within 20 kpc contains significant amount of angular momentum and is co-rotating with the cold disk.
The hot gas baryon fraction, $f_X=M_X/M_{\rm vir}=0.046$, is 3.8 times smaller than the cosmological baryon fraction. This 
implies an average density for the hot gas that is 3.8 times smaller than assumed in the standard ``cooling flow" halo model \`{a} la \citet{white91}, 
and yields a factor of 14.5 smaller X-ray emission measure. Contrary to the standard assumption that hot gas follows 
the radial distribution of the dark matter, the density distribution of hot gas in Eris follows a ``flattened" $\rho_X(r)\propto r^{-1.13}$ power-law 
profile out to 100 kpc (see Fig. \ref{fig5}). This gives origin to an X-ray surface brightness profile that is not as sharply peaked as
expected for hot halos with NFW profiles and that satisfies the observational constraints \citep[e.g.][]{anderson10}. The flattened 
density profile we find is consistent with the results of much lower-resolution simulations by \citet{crain10}, who identified 
the reason for the more extended gas distribution and weaker X-ray coronae in the entropy injection by SNe at $z\sim 1-3$. 
Only 10\% of the gas in Eris is in a very hot phase above $T=10^6$ K: 
the mean density of million degree gas at $R\ge 70$ kpc is $n\le 6\times 10^{-5}$ atoms cm$^{-3}$, which is well within the observational constraints 
from \OVI\ absorption measurements \citep{sembach03} in the halo of the Milky Way, and high enough to produce significant ram pressure stripping of 
dwarf spheroidal satellites \citep{mayer07}. 

The observed dispersion measure (DM) of pulsars in the Large Magellanic Cloud (LMC) provides another constraint to the hot halo of the
Milky Way \citep{anderson10}. Of the 11 pulsars discovered by \citet{manchester06} in the direction of the LMC, 3 have dispersion measures
below $45\,\dm$ and may be located within the Galaxy at a random position along the line of sight to the Cloud. The other 8
have dispersion measures in the range between 65 and 130 $\dm$ and are thought to be associated with the LMC,
with the lowest dispersion values belonging to pulsars located on the near side of the Cloud, some 50 kpc away. This leads to an estimate 
for the dispersion measure introduced by Galactic free electrons towards the LMC of ${\rm DM} \approx 70\,\dm$ \citep{anderson10}. To 
compare these values with prediction from the Eris simulation, we have calculated the 
mean integrated column density of free electrons between randomly positioned ``observers" on a circle in the stellar disk at a 
galactocentric distance of 8 kpc, and a ``pulsar" 50 kpc away. The mock pulsar was located at galactic coordinates $l=280^{\circ}$ and $b=-30^{\circ}$ 
like the LMC. The predicted dispersion measure,
\begin{equation}
{\rm DM} = \int_0^{50\,{\rm kpc}} n_e(l)dl=(62 \pm 3) ~{\rm cm^{-3}~pc}
\end{equation}
where $n_e$ is the electron density along the line of sight, is perfectly consistent with the data. Collectively, the above arguments
indicate that the reservoir of hot gas around the Eris simulated galaxy appears to satisfy pulsar dispersion-measure as well
as X-ray surface brightness and \OVI\ absorption constraints.

\begin{figure*}
\centering
\includegraphics[width=.47\textwidth]{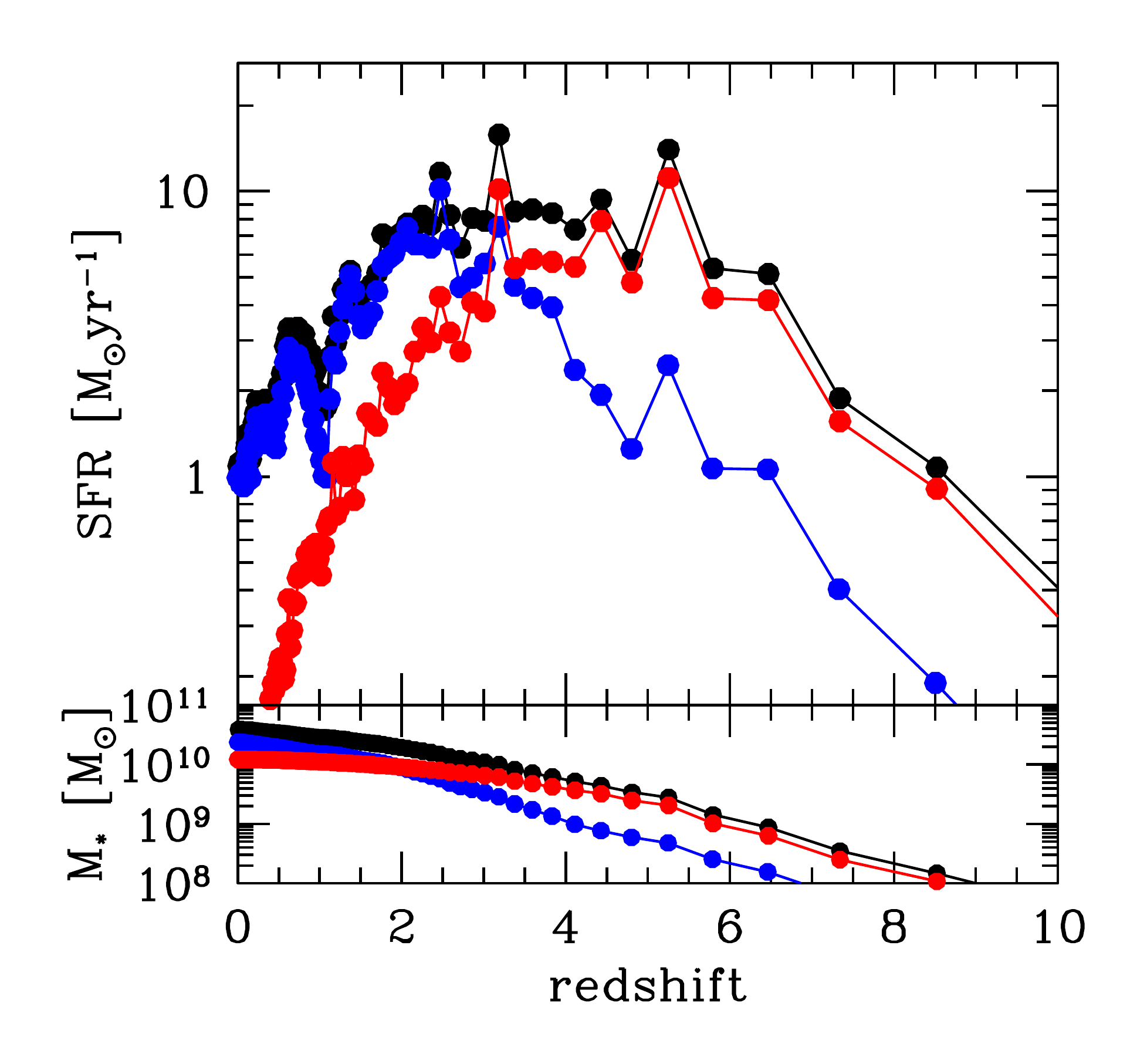}
\includegraphics[width=.47\textwidth]{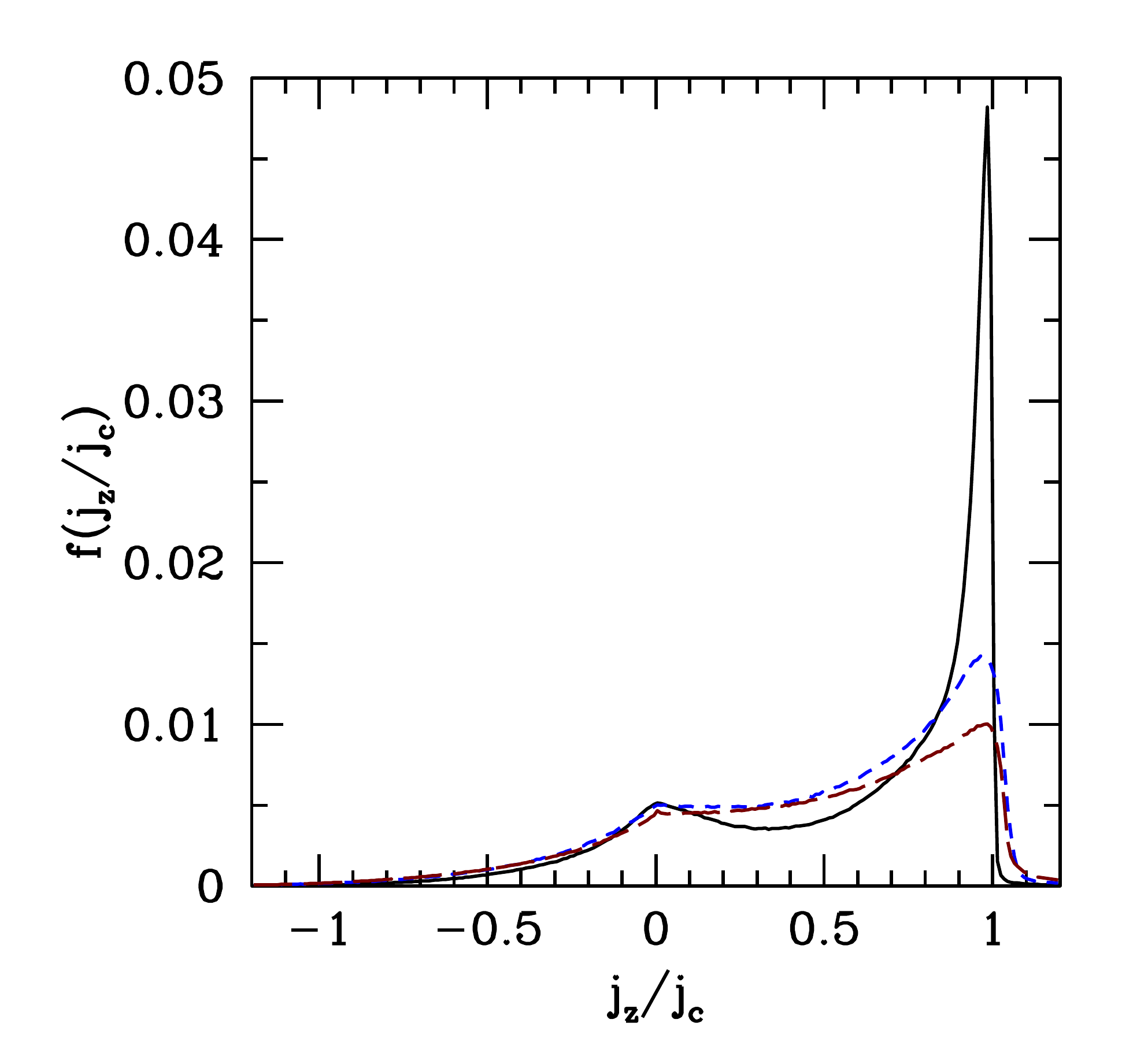}
\includegraphics[width=.47\textwidth]{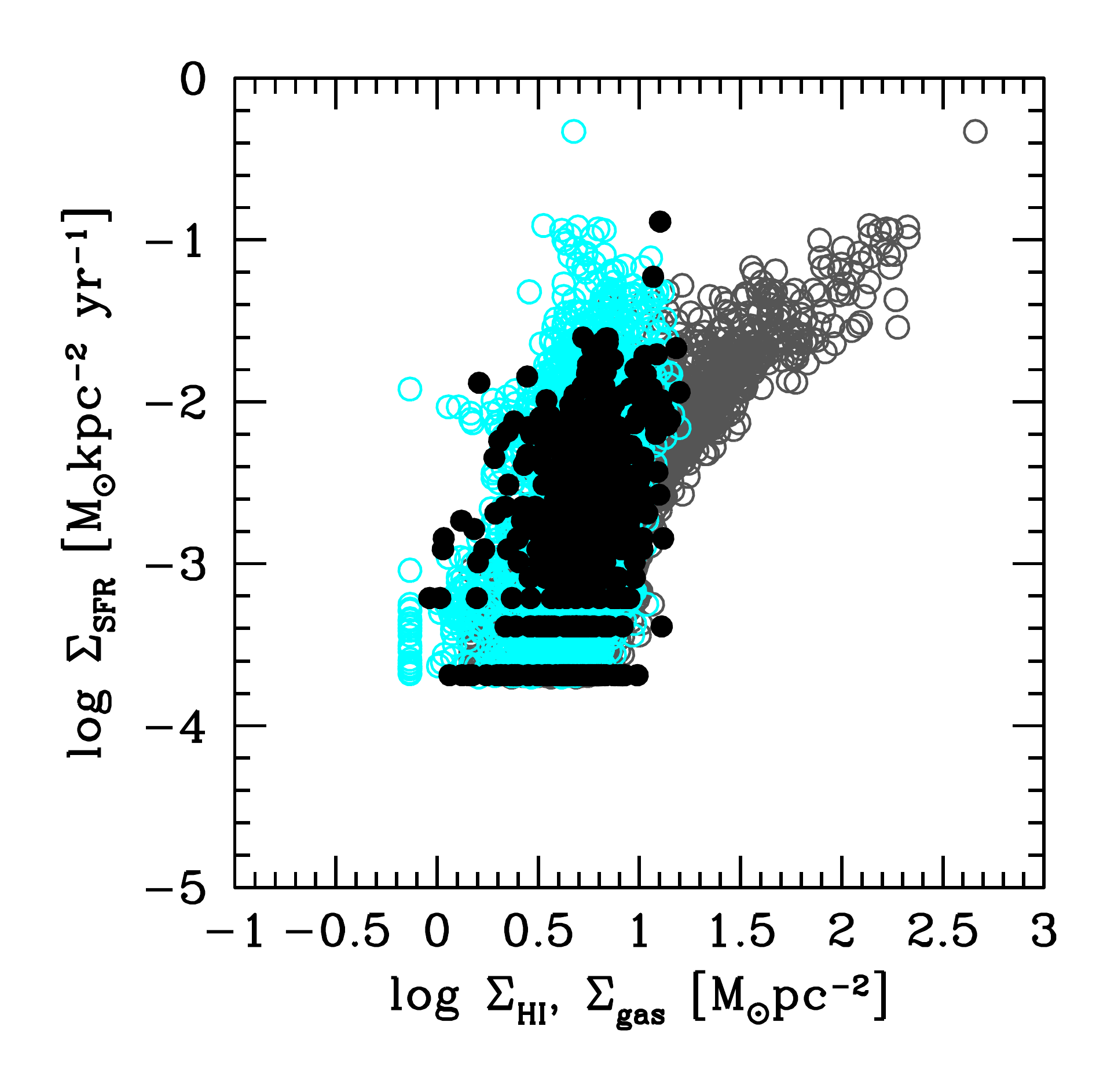}
\includegraphics[width=.47\textwidth]{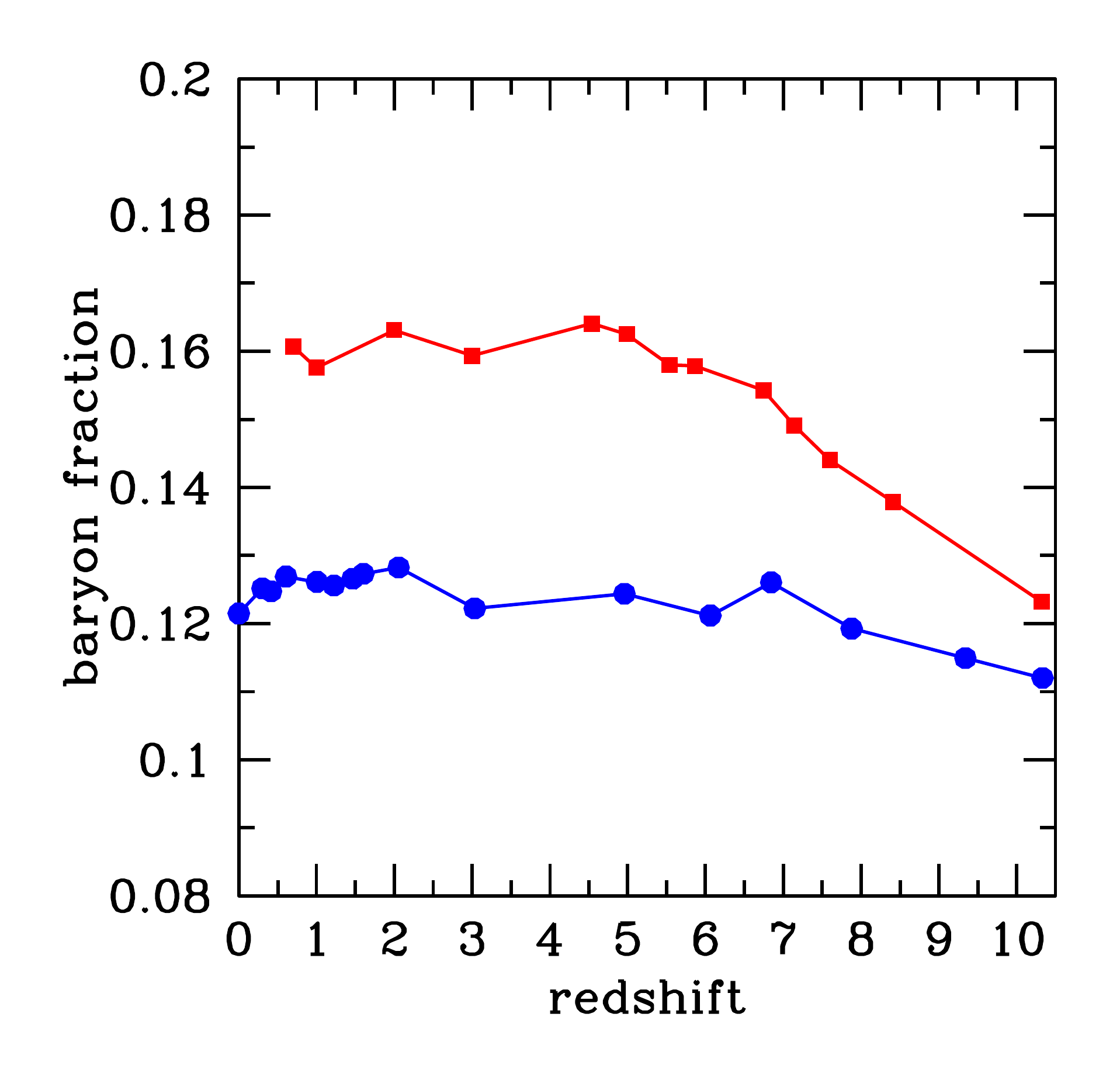}
\caption{{\it Top left:} Star formation history of all star particles identified within Eris's virial radius today. 
{\it Black filled dots}: total star formation rate ({\it top panel}) and stellar mass
({\it bottom panel}) as a function of redshift. {\it Blue filled dots}: same for disk star particles identified at $z=0$. 
{\it Red filled dots}: same for spheroid star particles identified at $z=0$. See the text for details of the disk-spheroid kinematic decomposition. 
{\it Top right:} Stellar mass fraction as a function of the ``orbital circularity parameter" $j_z/j_c$, describing the degree of 
rotational support of a given stellar particle, for Eris at $z=0$ ({\it solid line}), Eris at $z=1$ ({\it short-dashed blue line}), and
ErisLT at $z=1$ ({\it long-dashed red line}). The prevalence of stars in a centrifugally-supported thin disk manifests itself 
in a sharply peaked distribution about unity. {\it Bottom left:} $\Sigma_{\rm SFR}$ versus $\Sigma_{\rm HI}$ for Eris' disk at $z=0$ ({\it black filled 
dots}). The simulation data were averaged over square patches 750 pc on the side: some discreteness effects associated 
with the limited resolution of the star formation timescale can be seen at low values of $\Sigma_{\rm SFR}$. Every dot 
represents one sampling point. Note that our simulations do not model the formation of molecular hydrogen.
The {\it blue empty dots} show the pixel-by-pixel $\Sigma_{\rm SFR}$ data as a function of $\Sigma_{\rm HI}$ 
(at 750 pc resolution) for 7 spiral galaxies from the THINGS survey \citep{bigiel08}. The same THINGS data are plotted against the 
total gas surface density $\Sigma_{\rm gas}=\Sigma_{\rm HI}+\Sigma_{\rm H_2}$ ({\it gray empty dots}). 
{\it Bottom right:} Evolution of the baryon fraction within the virial radius for Eris ({\it blue filled dots}) and ErisLT 
({\it red filled squares}). In the adopted cosmology, the cosmic baryon fraction is $\Omega_b/\Omega_M=0.175$.}
\label{fig6}
\vspace{+0.3cm}
\end{figure*}

\subsection{Star Formation and Kinematic Decomposition}
The top left panel of Figure \ref{fig6} shows the star formation history of all star particles identified within Eris' virial radius at $z=0$
(regardless of whether they formed within the main host or in satellites), and of its kinematically-decomposed present-day {\it disk} and {\it 
spheroid}. The decomposition technique follows \citet{scannapieco09}, 
and is based on the distribution of orbital circularity parameters, $j_z/j_c$, of the simulated stars introduced by \citet{abadi03}.
Here, $j_z$ is the angular momentum of each star in 
the $z$-direction (i.e. the direction defined by the total angular momentum of all gas particles within 5 kpc from the host center) and $j_c$ is 
the angular momentum of a circular orbit at the same radius.
Spheroidal stars are defined as those that are not part of the rotationally-supported 
disk and therefore typically include bulge and stellar halo stars, as well as stellar bars if they are present.
The distribution of circularity parameters, shown in the top right panel of Figure \ref{fig6}, is characterized by two peaks: 
one at $j_z/j_c\simeq 1$ that is indicative of the presence of a dominant cold disk in rotational support, and a second one at $j_z/j_c\simeq 0$ 
corresponding to a modest hot spheroidal component dominated by velocity dispersion.

The vast majority of disk stars in the Milky Way reside in a thin disk component with exponential scale height $h_z=300\pm 60$ pc 
\citep{juric08}. A study of the vertical structure of edge-on spiral galaxies finds that the scale height of their thin disk 
increases systematically with circular velocity as $z_0=610$ $(V_c/100\,\kms)^{0.9}$ pc \citep{yoachim06}, where 
$z_0$ is the scale height of a sech$^2$ profile, $z_0\approx 2h_z$ at large heights above the disk plane.  
Eris' kinematically-decomposed stellar disk agrees well with the above scaling: by fitting an exponential (sech$^2$) profile to the 
simulation data, we derive a scale height of $h_z=490$ pc ($z_0=860$ pc) at a galactocentric distance of 8 kpc.

Today, Eris is forming stars at a rate of ${\rm SFR}= 1.1\,\sfr$, 
comparable to the value of ${\rm SFR} = 0.68-1.45\,\sfr$, recently inferred for the Milky Way by \citet{robitaille10} using {\it Spitzer} data. The star
formation rate declines rapidly with redshift from a plateau value of $\sim 10\,\sfr$ maintained between $z=2$ and $z=5$. 
SN feedback and photoheating by the ultraviolet radiation background efficiently quench star formation at $z>5$. 
The rate of formation of spheroidal stars fades rapidly after redshift 3, while disk stars nearly triple their mass from $z=2$ to the present. 

The star formation rate surface densities $\Sigma_{\rm SFR}$ and \HI\ gas surface densities $\Sigma_{\rm HI}$ (Kennicutt-Schmidt law) 
of Eris' disk are shown in the bottom-left panel of Figure \ref{fig6}. The simulation data were averaged over square patches 750 pc on the side for 
comparison with the pixel-to-pixel observations at 750 pc resolution of 7 spiral galaxies from the THINGS survey \citep{bigiel08}. 
The same THINGS data are also plotted against the total gas surface density $\Sigma_{\rm gas}=\Sigma_{\rm HI}+\Sigma_{\rm H_2}$.
The observed relationship between $\Sigma_{\rm gas}$ and $\Sigma_{\rm SFR}$ varies dramatically among and within spiral galaxies, and most 
galaxies show little or no correlation between $\Sigma_{\rm HI}$ and $\Sigma_{\rm SFR}$ \citep{bigiel08}. Rather, there is a clear 
correlation between $\Sigma_{\rm H_2}$ and $\Sigma_{\rm SFR}$ (molecular Schmidt law), and gas at densities in excess of $\sim 10 \,\msun$ pc$^{-2}$ 
is observed to be fully molecular. At solar metallicities, this corresponds to the column of atomic hydrogen required to shield a molecular region 
against photodissociation \citep{krumholz09}. While the total mass of cold gas in Eris is comparable to the sum of the atomic and molecular gas mass
of Sb galaxies such as the Milky Way, we do not model directly the formation of H$_2$ molecules. Indeed, as clearly shown in Figure 
\ref{fig6}, even at Eris's high resolution we do not capture kpc-sized regions with gas surface densities in excess of the characteristic shielding 
column. Furthermore, our gravitational softening 
is still large compared to the size of typical molecular clouds, which are a few tens of pc in size, and our gas density threshold for star
formation is still somewhat below the density of real molecular clouds. Both effects contribute to further 
smooth out the high density tail of the total gas distribution. Yet the figure shows that our simulated disk is forming stars in the same 
range of $\Sigma_{\rm SFR}-\Sigma_{\rm HI}$ values observed in spiral galaxies. 

The global atomic gas depletion timescale, i.e. the time needed for the present rate of star formation to consume the existing atomic gas reservoir,
is $t_{\rm HI}\equiv M_{\rm HI}/{\rm SFR}=1.9\times 10^9\,\msun/1.1\,\sfr=1.7$ Gyr in Eris today. As this is significantly smaller than the Hubble time, 
star formation rates and gas fractions must be set by the balance between gas accretion from the halo and stellar feedback. The depletion 
time for atomic gas observed in the COLD GASS survey \citep{saintonge11} is on average 3 Gyr, with a large scatter from one galaxy to another.  

\section{Discussion}

Approximately 70\% of bright spirals have $B/T\le 0.2$ \citep{weinzirl09}. Recent attempts to generate such disk-dominated galaxies in cosmological 
simulations have failed to reproduce simultaneously their observed morphologies as well as their baryonic/stellar content \citep{scannapieco10,agertz11}. 
Our very high-resolution Eris simulation appears to form a close analog of a Milky Way disk galaxy by capturing a realistic 
inhomogeneous ISM in which star formation occurs in high density regions of mass comparable to that of giant cloud complexes. 
Contrary to the ``inefficient star formation'' prescription adopted by \citet{agertz11}, this is achieved with a strong localized SN feedback and 
a high (10\%) Schmidt-law efficiency.  We stress that our efficiency parameter is simply phenomenological, and does not
have a direct relation to the true star formation efficiency within giant molecular clouds, which is the end result of 
all the processes regulating star formation including feedback, and concerns scales (tens of parsecs) that are not yet resolved in 
cosmological simulations. While the chosen star formation and feedback prescriptions combine to expel a significant amount of 
baryons from the system, they preferentially remove low angular momentum material and leave plenty of cold gas available for disk star
formation \citep[cf.][]{scannapieco09}, thus allowing a good fit to the Tully-Fisher relation \citep[cf.][]{piontek11}. 

It is interesting to compare the properties of Eris and ErisLT at redshift 1 (see Table 1):

\begin{itemize}

\item The ErisLT control run produces a galaxy resembling an early-type Sa spiral with $B/D=0.42$ (cf. Eris' $B/D=0.31$), closer to 
(but still on the low side of) the typical outcome of previously published cosmological simulations of disk formation 
\citep[e.g.][]{governato10,scannapieco10}. Its rotation curve peaks at $308\,\kms$ (cf. Eris' $237\,\kms$) and declines steeply within 
few kpc from the center (Fig. \ref{fig1}). 

\item  
The baryon fraction in ErisLT is higher than in Eris and close to the universal value. The difference in the baryon content
of the two simulations is established at very early times (see the bottom right panel of Fig. \ref{fig6}). The baryon content of ErisLT 
increases from $z=10$ to $z=5$, as its dark matter halo grows from $M_{\rm vir} = 3\times 10^9\,\msun$ to $M_{\rm vir} = 6\times 10^{10}\msun$. 
At higher redshifts (smaller progenitor mass), the collapse of baryons is heavily suppressed by the ultraviolet radiation background. Eris, by 
contrast, maintains a relatively low baryon fraction, between 64\% and 73\% of the cosmic value at all redshift $z<10$. 
Note that these values are higher than the value measured at the present epoch in the bulgeless dwarf galaxy simulation of \citet{governato10},   
which is only 30\% of the cosmic fraction.

\item 
While ErisLT was run with a star formation efficiency that was half of that adopted for Eris, its stellar content at $z=1$ is 20\% 
higher, showing that the most important parameter in the star formation recipe is not the efficiency but rather the star formation 
density threshold. Indeed, by allowing the gas to reach higher densities before turning into stars, 
the ISM develops a more inhomogeneous structure, with important consequences on the large-scale pattern of star formation in 
the galaxy, and, as a byproduct, on the effect of supernovae feedback \citep[e.g.][]{governato10}.

\begin{figure*}
\centering
\includegraphics[width=.47\textwidth]{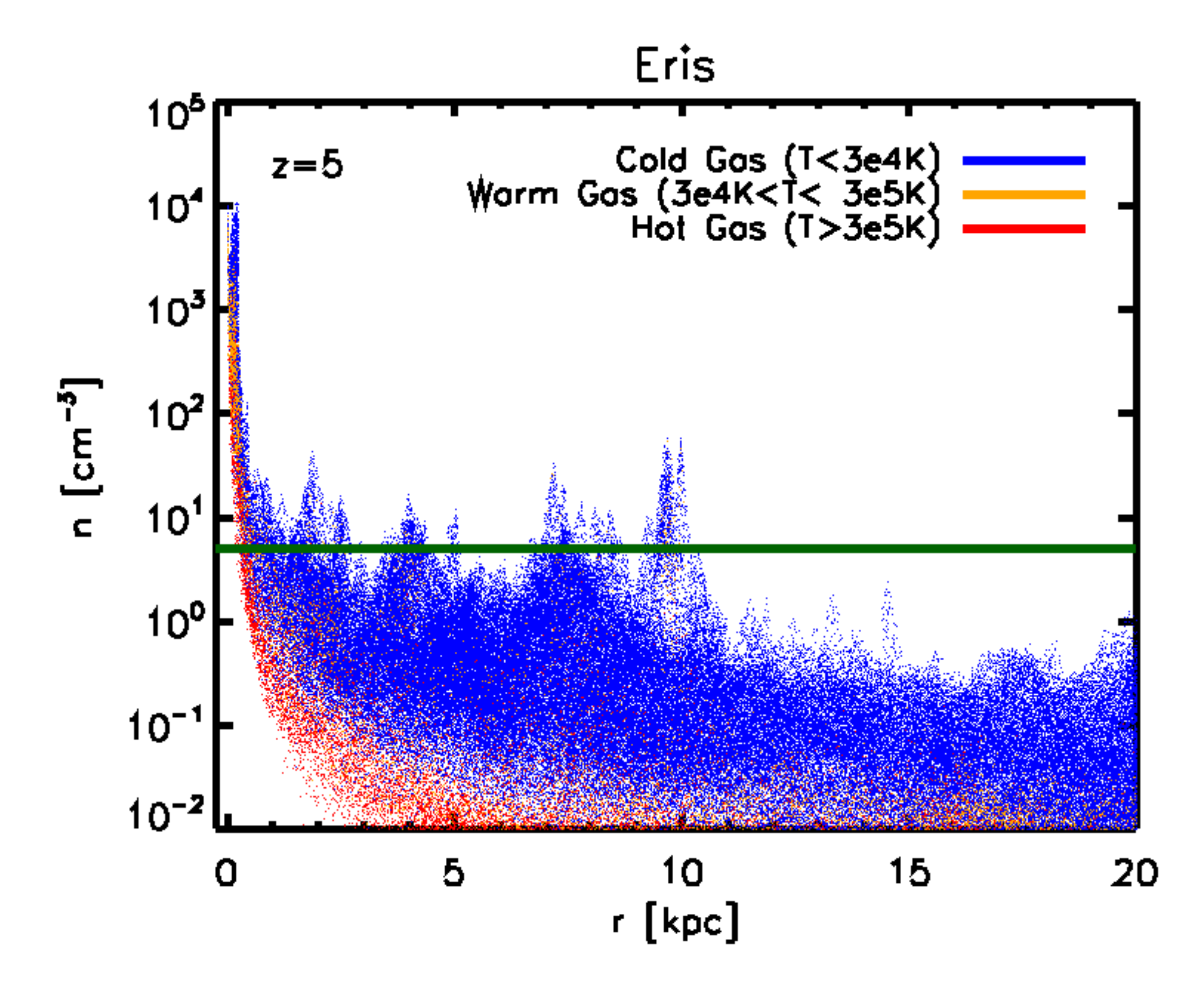}
\includegraphics[width=.47\textwidth]{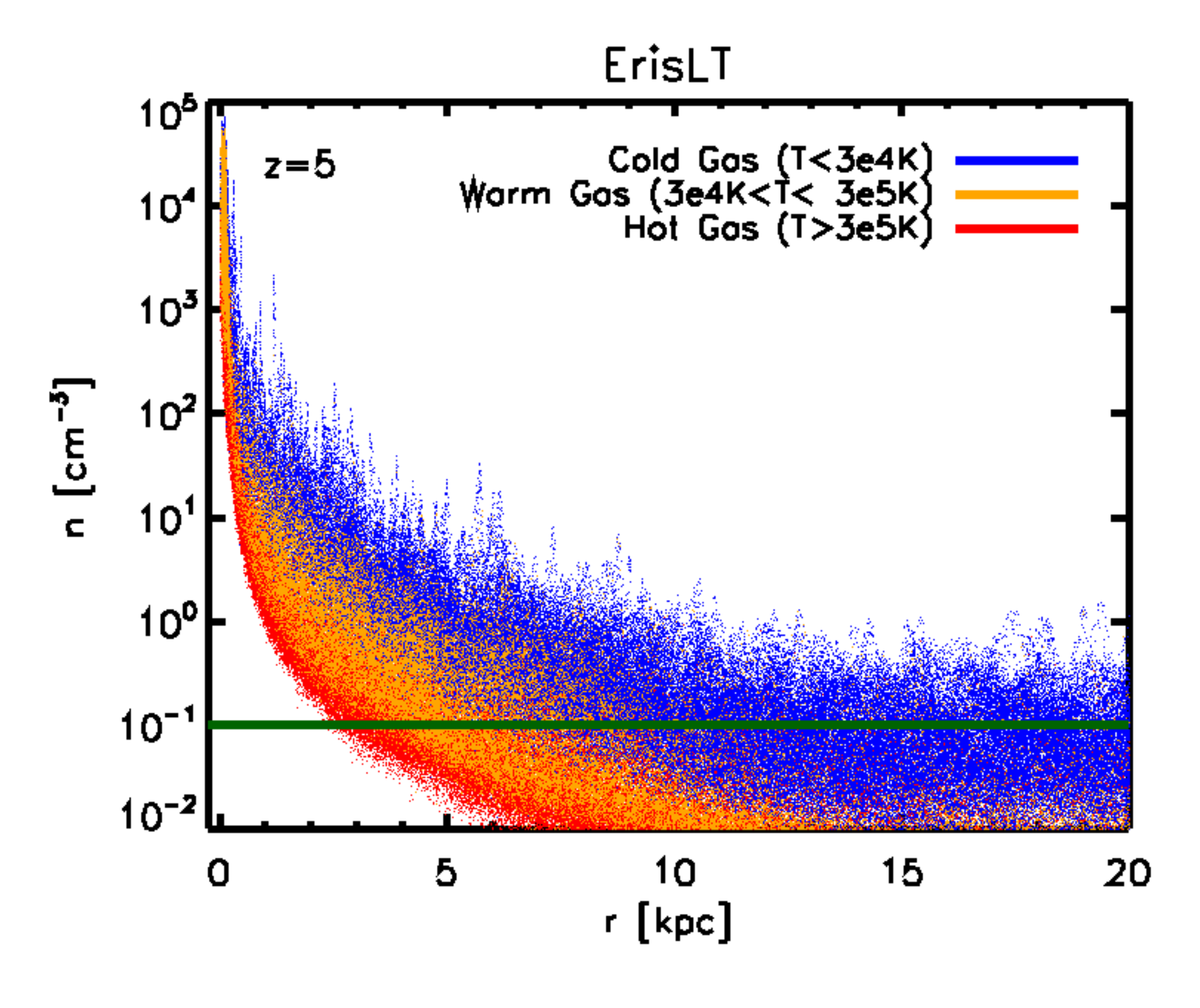}
\includegraphics[width=.47\textwidth]{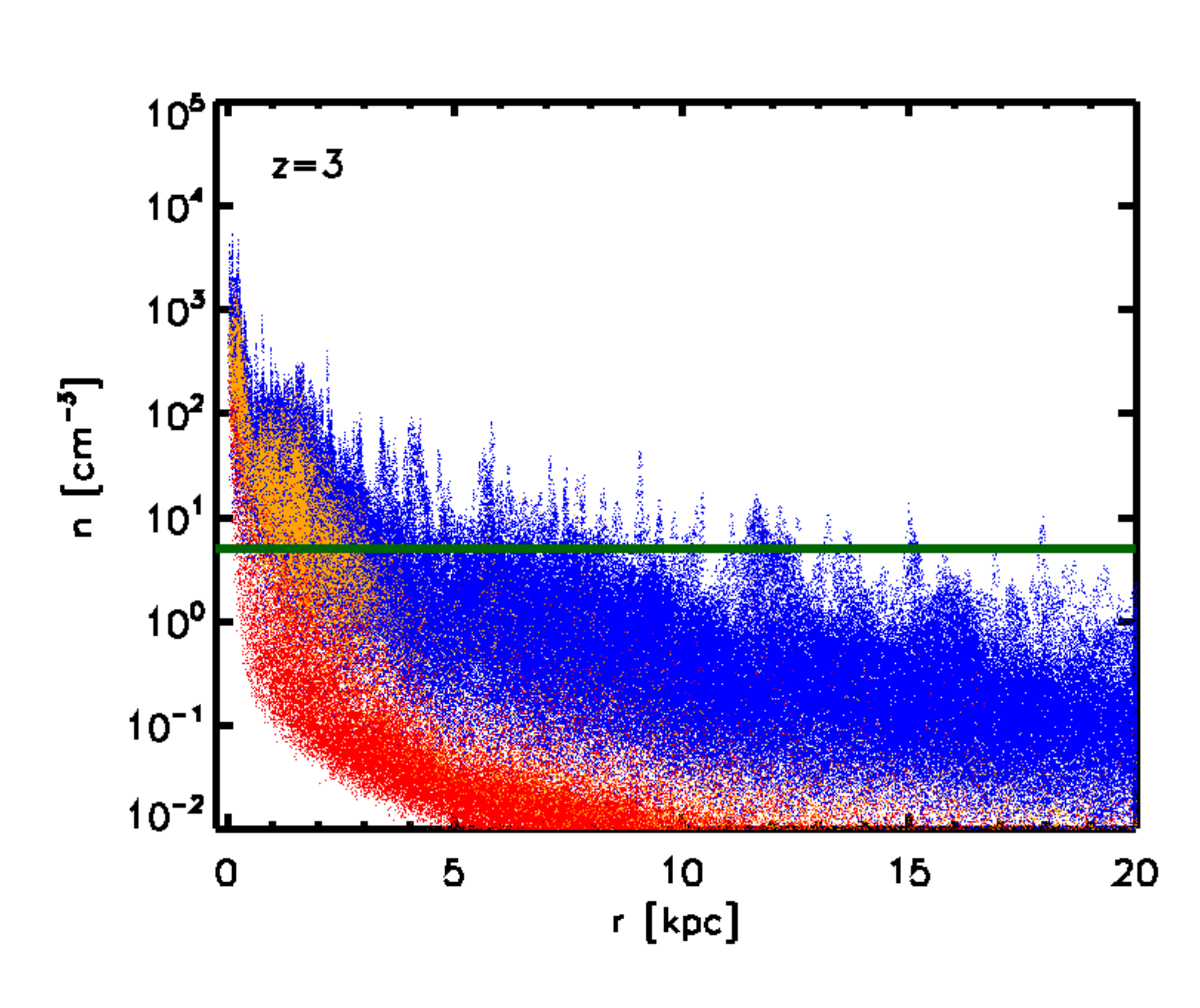}
\includegraphics[width=.47\textwidth]{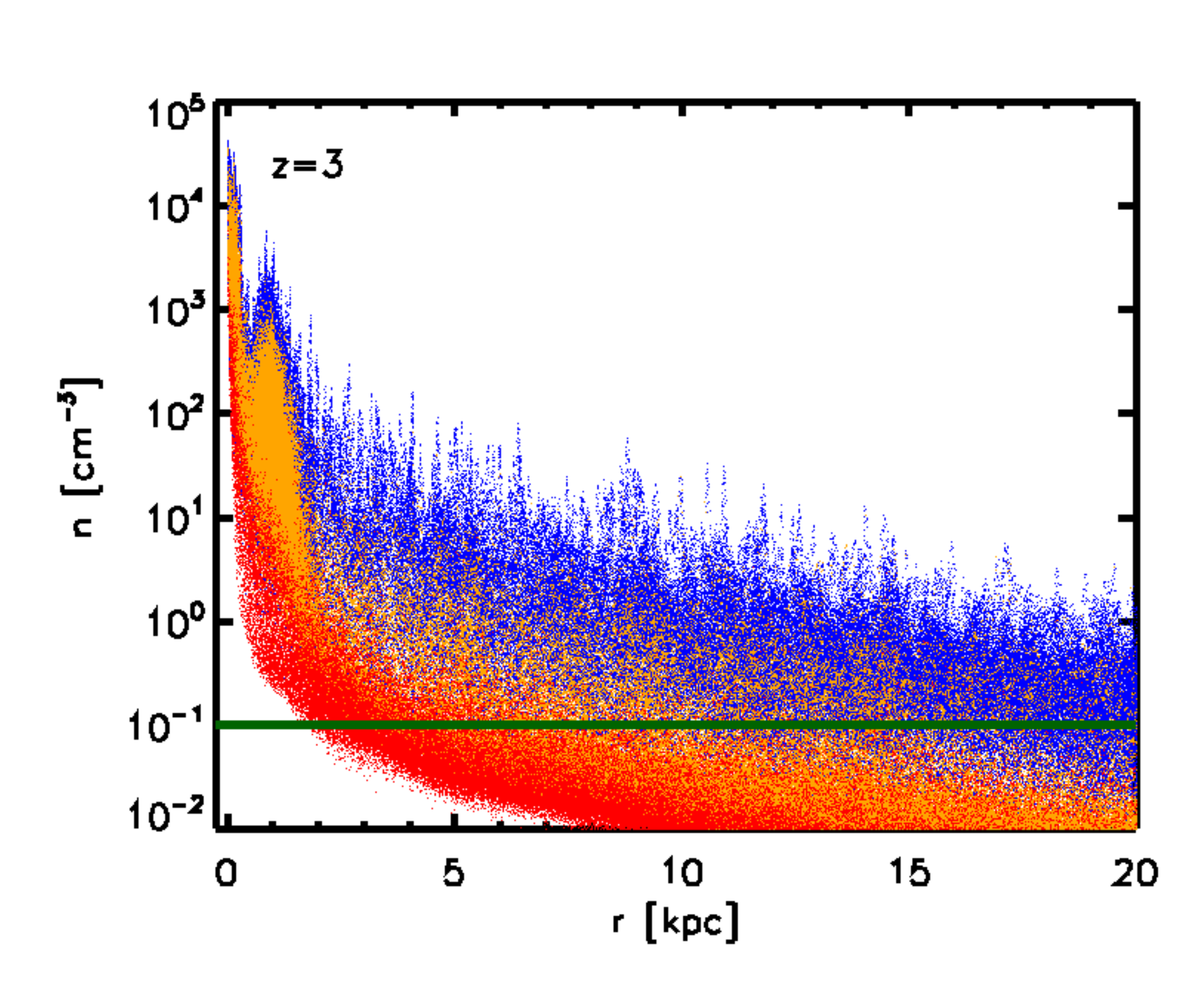}
\includegraphics[width=.47\textwidth]{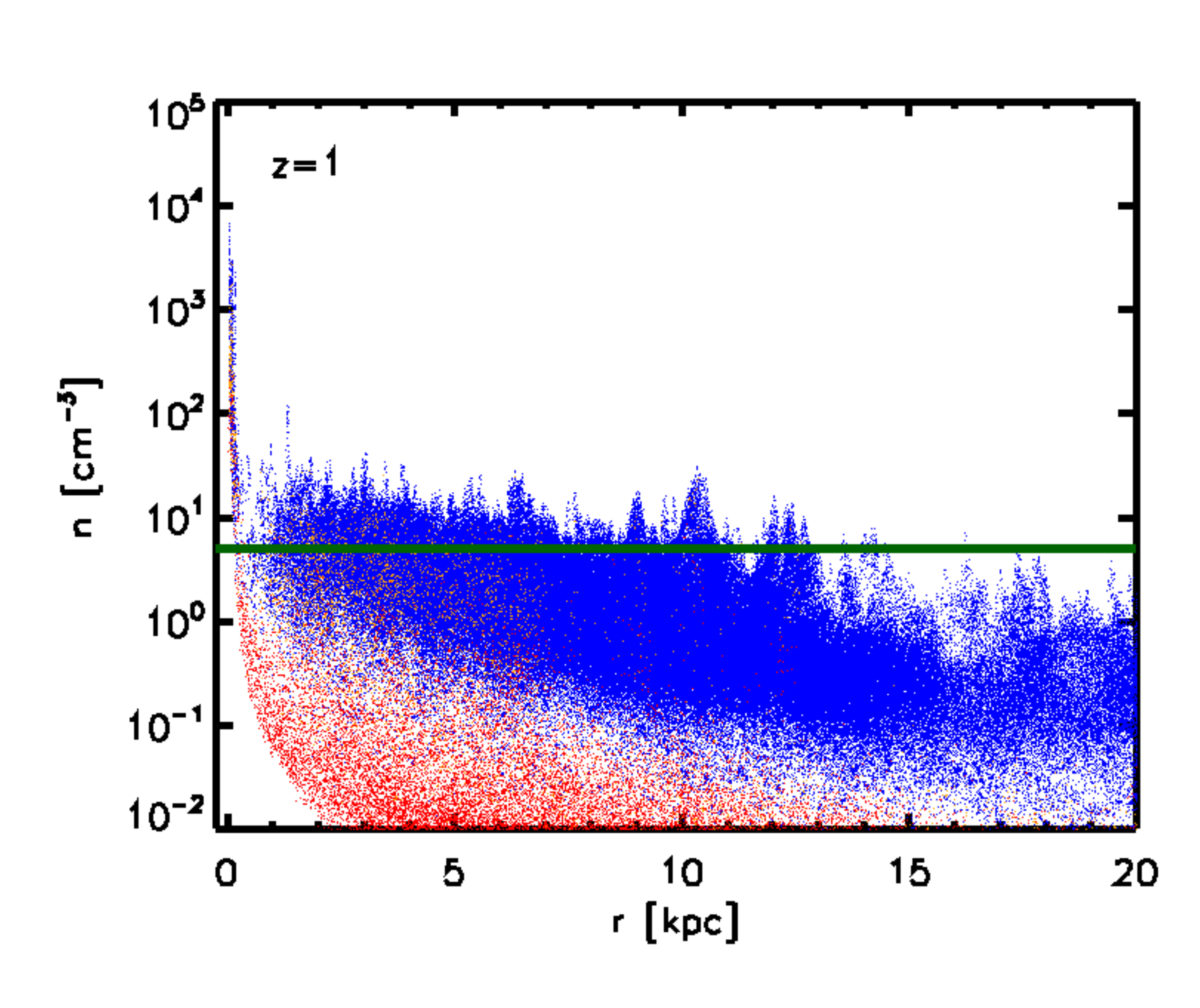}
\includegraphics[width=.47\textwidth]{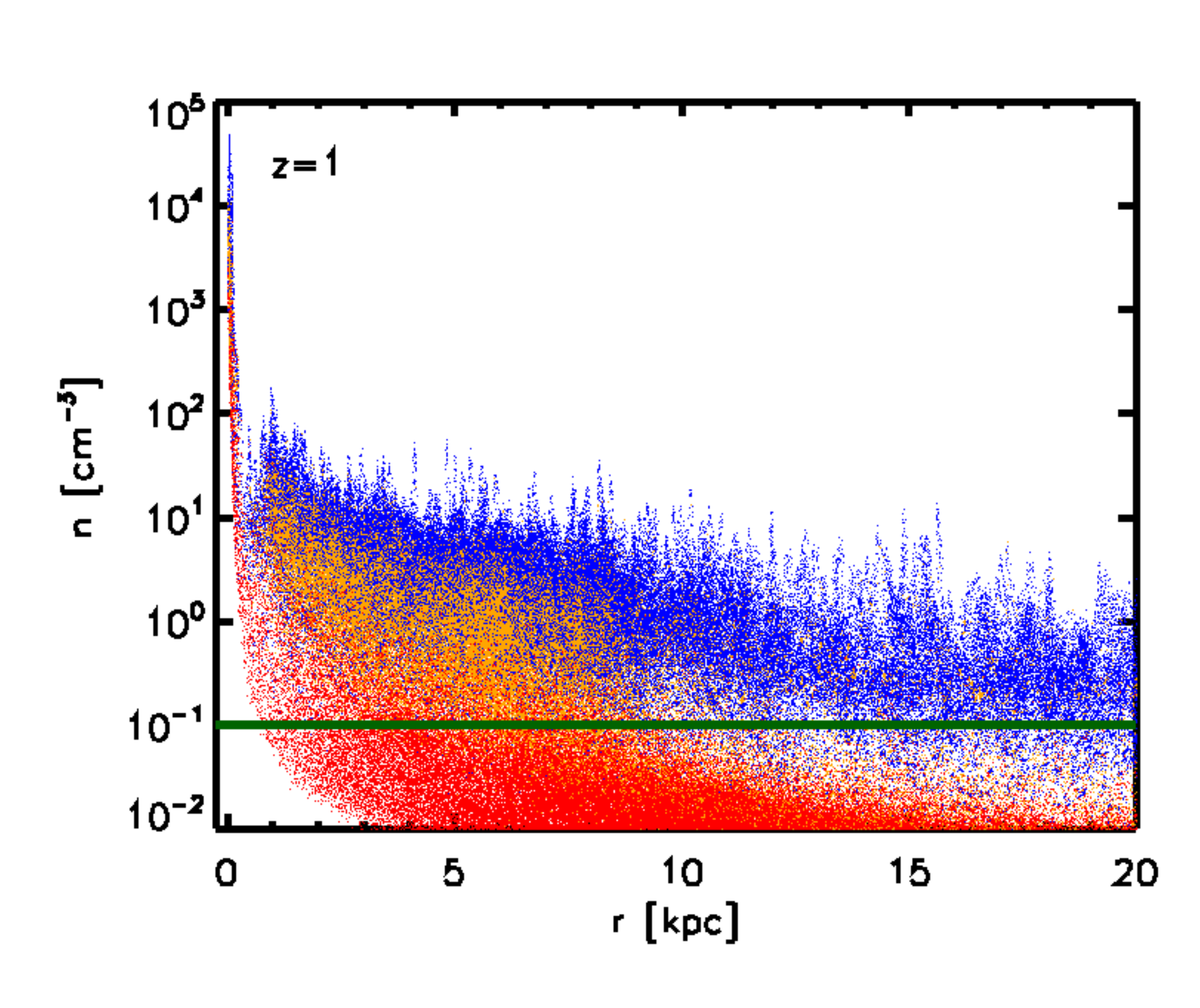}
\caption{Properties of the gas distribution for different star formation thresholds and different redshifts. The local gas density
measured around each SPH particle is plotted as a function of its radial distance from the galaxy center for Eris ({\it left panels})
and ErisLT ({\it right panels}) at $z =$ 1, 3, and 5 (from bottom to top). Horizontal green lines mark the minimum gas density for star 
formation in each run.
The color coding highlights cold ($T<3\times 10^4$ K, {\it blue}), warm ($3\times 10^4\,{\rm K}<T<3\times 10^5$ K, {\it yellow}), and hot 
($T>3\times10^5$ K, {\it red})) gas particles. Star formation only occurs in blue gas particles above the horizontal green lines.    
}
\label{fig7}
\vspace{+0.cm}
\end{figure*}

\item 
At $z=1$, about 14\% of the total gas content in ErisLT is in a cold phase, 19\% is warm, and 67\% is in a hot component. 
The same fractions are 30\%, 18\%, and 52\% in Eris, i.e. there is 2 times more cold gas in Eris than in ErisLT. 
The difference in the density distribution and thermodynamic state of the ISM in the two simulations is clearly seen in Figure \ref{fig7},
which shows the local gas density versus radius of each gas particle at different epochs. The horizontal green line in each panel marks
the adopted star formation threshold, and particles are color coded according to their temperature. Stars are born only in the blue gas particles above the 
green lines. Because of the increased threshold, fewer regions are forming stars at any given time in Eris. According to equation (\ref{eq:KS}), however, 
within these cold dense clumps the number of massive stars per unit gas mass born at threshold scales as $\sqrt{n_{\rm SF}}$, i.e. it is a 
factor $(5/0.1)^{1/2}=7$ times larger in Eris versus ErisLT. As these stars explode as SNe, the energy injected at high redshift in Eris's ISM 
is enough to disrupt and unbind the star forming regions, generating strong outflows that leave the galaxy and reduce its baryonic content. 

\item 
Figure \ref{fig7} also shows that, at $z=5$, star formation can only occur in Eris' inner densest regions within 10 kpc from the center. By contrast, in ErisLT, 
star formation is spread out over a significant fraction of the host galaxy, extending well beyond 20 kpc. At $z=3$, when outflows are stronger 
due the high star formation rates triggered by mergers, 
it is the cold, low angular momentum gas at the center that is preferentially removed in Eris, as previously found for dwarf galaxies by 
\citet{governato10}. Such outflows  are weaker in the low threshold simulation, where star formation is more diffuse and the energy injection 
from SNe is more evenly deposited in the ISM: cold gas accumulates in the inner regions already at very high redshifts and continues to turn 
into stars unabated by feedback \citep[see also][]{ceverino09,robertson08}. At lower redshift ErisLT has now consumed its low-angular momentum cold 
gas in star formation, while Eris has preserved a higher cold gas fraction due to the more effective regulation of star formation via feedback.
By $z=1$ both galaxies have formed roughly the same amount of stars, but 90\% of the gas within 20 kpc from the center is still cold in Eris, compared 
to only 57\% in ErisLT. The distributions of stellar angular momenta are also different in the two cases (see Fig. \ref{fig6}).  
Gas being blown out at high redshift has a systematically lower angular momentum than the gas that 
gets accreted at later times \citep{brook10}, and this explains the flatter rotation curve of Eris relative to ErisLT. 

\end{itemize}

It is fair to point out that, while the success of our Eris simulation in matching the observations appears to be linked to the ability   
of correctly following star formation in an inhomogenous ISM and regulating it with SN feedback, many avenues remain unexplored and require further 
investigation. Simulations of even higher resolution, approaching the true gas densities reached in star forming giant molecular cloud
complexes (about a factor of 10 higher than adopted here), are needed to test the convergence and robustness of our results, and
are in the making. The density threshold should depend on metallicity and therefore on redshift \citep{gnedin09}, which may affect the structure 
of the ISM and SN feedback differently in progenitors at different epochs. The feedback model that
we use is still phenomenological, and the actual mechanism of outflow generation may require the combination of
more than one effect to support a large-scale blastwave \citep[e.g.][]{ceverino09}.  Lastly, the simulations presented in this paper 
neglect cooling by metal lines at temperatures above $10^4$ K. At the mass scale considered here, cold flows are mostly responsible for the
assembly of the star forming disk even at low redshift, as opposed to the cooling flow of the hot halo mode \citep{brooks09}.
This suggests that the details of the cooling function for gas above $10^4$ K, namely in the hot mode, are not important for the assembly of the disk. 
\citet{piontek11} find that metal cooling gives origin to a stronger burst of star formation at high redshift in the progenitors of 
Milky-Way sized  galaxies and thus to bigger bulges. This result, however, was obtained with the canonical low star formation density threshold 
($n_{\rm SF}=0.1$ atoms cm$^{-3}$), and it is unclear in which temperature range the effect of metal cooling is most crucial. These authors also show that 
the augmented bulge can be suppressed by boosting the effect of (thermal and kinetic) SN feedback. It is conceivable that, by further increasing the
density threshold for star formation toward actual giant molecular cloud densities, along with modeling the formation of molecular hydrogen,
the ISM may become increasingly clumpy and dense, populating the high surface density tail of the Kennicutt-Schmidt relation that remains 
currently unresolved in Eris. As a result, one may expect that heating and outflows from even more localized SN feedback may become stronger in 
such high density regions, and eventually offset the impact of increased cooling, as suggested by \citet{piontek11}. We plan to explore these issues
in the next generation of simulations with increased resolution, a higher star formation density threshold, and the inclusion of molecular phase physics. 
In addition to the effect on disk and bulge, metal cooling at $T>10^4$ K may also have an impact on the density profile and temperature of the 
hot halo. \citet{piontek11} find a significant reduction of the mass of the hot halo when metal cooling is included, under the assumption 
of a gas metallicity of $Z = 0.5 Z_{\odot}$. This is higher than the metallicity measured in spiral galaxies with extended X-ray and H$\alpha$ emission 
around their disk, which is in the range $Z= 0.01-0.1 Z_{\odot}$ \citep{rasmussen09}, suggesting that the effect may have been overestimated. 
The mean metallicity of hot gas in Eris closely matches the latter observations, being on average $Z=0.08 Z_{\odot}$ (with $Z_\odot=0.0194$, \citealt{anders89}).

The last major merger in our simulations occurs at $z\sim3$, and therefore Eris is expected to show some offset from the observed 
structural parameters of the average spiral galaxy. The same likely applies to the Milky Way itself, which indeed closely resembles Eris. 
Whether or not the good match with the properties of typical spiral galaxies is related to the fact that we have selected a particularly quiet
merging history, in which outflows shut off early as the rate of star formation drops after the last major merger, will have
to be investigated. 

At $z<3$, the evolution of the B/D ratio is non-monotonic and it is seen to decrease following a major merger at $z=3$ and 
a minor merger at $z=1$, and to grow secularly at lower redshifts. The details of this evolution will be the subject of a forthcoming paper (Guedes et al. 2011c, in preparation).
More typical galaxies undergoing significant mergers at later times may 
develop smaller bulges due to the more prolonged effect of supernovae outflows, which could explain why 11 out of 19 nearby massive spirals 
show no evidence for a classical bulge \citep{kormendy10}). If true, the trend
would be at odds with the standard picture in which mergers lead to earlier type objects. This would be an important new ingredient, perhaps 
complementary to the finding that gas-rich mergers assist the growth of larger disks \citep{hopkins09,governato09}. 

\acknowledgements

This research was funded by NASA through grant NNX09AJ34G and by the NSF through grant AST-0908910 (PM), by the Swiss National Foundation 
(SNF), and by an ARCS Foundation Fellowship to JG. Simulations were carried out on NASA's Pleiades supercomputer, the UCSC Pleiades cluster, 
and the Swiss National Supercomputing Center's ROSA Cray-XT5. We thank the referee for useful comments that improved this paper and acknowledge 
useful discussions on the topic of this paper with Oscar Agertz, Alyson Brooks, Valentino Gonz\'alez, Fabio Governato, Du\v{s}an Kere\v{s}, 
Andrey Kravtsov, Mark Krumholz, Brant Robertson, Sijing Shen, and Romain Teyssier. We are indebted to Frank Bigiel for helping with THINGS data, 
Charlie Conroy for providing a photometric stellar mass estimate for Eris, and Elena D'Onghia for helping in generating the initial conditions and selecting the Eris halo. LM thanks the Aspen Center of Physics for hospitality during the early stages of the work.

\end{document}